{\global\def\bigPage{
        \setlength{\topmargin}{-.75in}
        \setlength{\textheight}{9.5in}
        \setlength{\oddsidemargin}{+0.0in}
        \setlength{\textwidth}{6.2in}
        }
}

\documentclass[11pt,onecolumn,draftcls]{IEEEtran}\normalsize
\usepackage{amsfonts,amsthm,amsmath,epsf,amssymb,cite}
\usepackage[usenames,dvipsnames]{color}
\usepackage[mathscr]{eucal}

\bigPage
\newcommand{\defn}{\ensuremath{\triangleq}}  
\newcommand{\limit}[2]{\ensuremath{\lim_{#1 \rightarrow #2}}}  
\newcommand{\tvec}[1]{\ensuremath{\Tilde{\boldsymbol{#1}}}}

\newcommand{\hvec}[1]{\ensuremath{\Hat{\boldsymbol{#1}}}}
   
\renewcommand{\vec}[1]{\ensuremath{\boldsymbol{#1}}}

\newcommand{\norm}[1]{\ensuremath{\| #1 \|}}  
\newcommand{\mc}[1]{\ensuremath{\mathcal{#1}}}  
\newcommand{\st}{{~\text{s.t.}~}}

\newcommand{\Complex}{{\mathbb{C}}}

\newcommand{\conv}{\ast}

\newcommand{\tran}{^\text{\sf T}}
\newcommand{\herm}{^\text{\sf H}}
\newcommand{\pherm}{^\text{\sf +H}}
\newcommand{\of}[1]{^{(#1)}}
\newcommand{\ofc}[1]{^{{(#1)}*}}  
\newcommand{\oft}[1]{^{{(#1)}\text{\sf T}}}  
\newcommand{\ofH}[1]{^{{(#1)}\text{\sf H}}}
\newcommand{\ofP}[1]{^{{(#1)}\perp}}

\DeclareMathOperator{\E}{E}  
  
\DeclareMathOperator{\cov}{cov}

\DeclareMathOperator{\tr}{tr}  
\DeclareMathOperator{\Diag}{\mc{D}}  
\DeclareMathOperator{\diag}{diag}

\DeclareMathOperator*{\argmin}{arg\,min}
\DeclareMathOperator*{\argmax}{arg\,max}
\DeclareMathOperator{\mi}{I}
\DeclareMathOperator{\en}{h}

\newtheorem{theorem}{Theorem}  
\newtheorem{lemma}{Lemma}  
  
\newtheorem{corollary}{Corollary}

\renewcommand{\eqref}[1]{(\ref{eq:#1})}

\newcommand{\tabref}[1]{Table~\ref{tab:#1}}  
\newcommand{\secref}[1]{Section~\ref{sec:#1}}  
\newcommand{\appref}[1]{Appendix~\ref{app:#1}}  

\newcommand{\lemref}[1]{Lemma~\ref{lem:#1}}  
\newcommand{\thmref}[1]{Theorem~\ref{thm:#1}}  
\newcommand{\corref}[1]{Corollary~\ref{cor:#1}}

\newcommand{\textb}[1]{#1}

\newcommand{\ccs}{_{\text{\sf ccs}}}

\newcommand{\nonsparse}{_{\text{\sf non-sparse}}}
\newcommand{\sparse}{_{\text{\sf sparse}}}
\newcommand{\f}{_{\text{\sf f}}}
\newcommand{\ik}{_{i_k}}
\newcommand{\ikt}{_{i_{k,\text{\sf true}}}}

\newcommand{\fpi}{_{\text{\sf f,p},i}}
 \newcommand{\fpik}{_{\text{\sf f,p},i_k}}
 \newcommand{\fpikt}{_{\text{\sf f,p},i_{k,\text{\sf true}}}}

\newcommand{\nz}{_{\text{\sf nz}}}
 \newcommand{\nzp}{_{\text{\sf nz,p}}}
 \newcommand{\nzi}{_{\text{\sf nz},i}}
 
 \newcommand{\nzpi}{_{\text{\sf nz,p},i}}
 \newcommand{\nzpik}{_{\text{\sf nz,p},i_k}}
 
\newcommand{\nzt}{_{\text{\sf nz,p,true}}}
\newcommand{\pl}{_{\text{\sf p}}}
\newcommand{\pli}{_{\text{\sf p,}i}}
\newcommand{\plj}{_{\text{\sf p,}j}}
 
\newcommand{\plt}{_{\text{\sf p,true}}}
\newcommand{\da}{_{\text{\sf d}}}
\newcommand{\dat}{_{\text{\sf d,true}}}
\newcommand{\di}{_{\text{\sf d},i}}
\newcommand{\dii}{_{\text{\sf d},\vec{i}}}
 \newcommand{\dik}{_{\text{\sf d},i_k}}
 \newcommand{\dikt}{_{\text{\sf d},i_{k,\text{\sf true}}}}
\newcommand{\true}{_{\text{\sf true}}}
\newcommand{\ml}{^{\text{\sf ML}}}
\newcommand{\map}{^{\text{\sf MAP}}}
\newcommand{\wmd}{^{\text{\sf WMD}}}
\newcommand{\giv}{\,|\,}
\newcommand{\biggiv}{\,\big|\,}
\newcommand{\Biggiv}{\,\Big|\,}
\newcommand{\code}{\mathfrak{C}}

\newcommand{\dasd}{^{\text{\sf DASD}}}
\newcommand{\ii}{_{\vec{i}}}
\newcommand{\iip}{_{\vec{i}'}}
\newcommand{\iis}{_{\vec{i}_{\text{\sf stop}}}}
\newcommand{\iit}{_{\vec{i}_{\text{\sf true}}}}

\newcommand{\last}{_{\text{\sf last}}}
\newcommand{\stp}{_{\text{\sf stop}}}
\newcommand{\pasd}{^{\text{\sf PASD}}}

\begin{document}

\title{On Communication over Unknown Sparse Frequency-Selective Block-Fading Channels}

\author{Arun~Pachai~Kannu and Philip~Schniter,~\IEEEmembership{Senior Member,~IEEE}%
	\thanks{
	Arun Pachai Kannu is with the Dept.\ of Electrical Engineering at the
	Indian Institute of Technology, Madras, and 
	Philip Schniter is with the Dept.\ of Electrical and Computer 
	Engineering at The Ohio State University, Columbus OH, 43210.
	Please direct all correspondence to Philip Schniter, Dept.\ ECE,
	2015 Neil Ave., Columbus OH 43210, e-mail schniter@ece.osu.edu,
	phone 614-247-6488, fax 614-292-7596.
	This work was supported in part by NSF grant CCF-1018368
	and DARPA/ONR grant N66001-10-1-4090.
	}
	}

\maketitle

\vspace{-14mm}

\begin{abstract}
This paper considers the problem of reliable communication over 
discrete-time channels whose impulse responses have length $L$ and 
exactly $S\leq L$ non-zero coefficients, and whose support and coefficients 
remain fixed over blocks of $N>L$ channel uses but change independently 
from block to block. 
Here, it is assumed that the channel's support and coefficient realizations 
are both unknown, although their statistics are known.
Assuming Gaussian non-zero-coefficients and noise, and focusing on the
high-SNR regime, it is first shown that 
the ergodic noncoherent channel capacity 
has pre-log factor $1-\frac{S}{N}$ for any $L$.
It is then shown that, to communicate with arbitrarily small error 
probability at rates in accordance with the capacity pre-log factor, 
it suffices to use pilot-aided 
orthogonal frequency-division multiplexing (OFDM)
with $S$ pilots per fading block, in conjunction with an
appropriate noncoherent decoder.
Since the achievability result is proven using a noncoherent decoder whose 
complexity grows exponentially in the number of fading blocks $K$, 
a simpler decoder, based on $S+1$ pilots, is also proposed. 
Its $\epsilon$-achievable rate is shown to have
pre-log factor equal to $1-\frac{S+1}{N}$ with the previously considered 
channel, while its achievable rate is shown to have pre-log factor 
$1-\frac{S+1}{N}$
when the support of the block-fading channel remains fixed over time.
\end{abstract}

\begin{IEEEkeywords}
Bayes model averaging, 
compressed sensing, 
fading channels,
noncoherent capacity, 
noncoherent communication, 
sparse channels. 
\end{IEEEkeywords}

\section{Introduction}					\label{sec:intro}

We consider the problem of communicating reliably over an \emph{unknown sparse} 
single-input single-output (SISO) frequency-selective block-fading 
channel that is described by the discrete-time complex-baseband 
input/output model
\begin{eqnarray}
 y\of{k}[n] 
 &=& \sqrt{\rho} \sum_{l=0}^{L-1} h\of{k}[l] x\of{k}[n-l] + v\of{k}[n],
							\label{eq:ykn}
\end{eqnarray}
where 
$n \in \{0,\dots,N-1\}$ is the channel-use index, 
$k \in \{1,\dots,K\}$ in the fading-block index,
$x\of{k}[n]$ is the transmitted signal,
$y\of{k}[n]$ is the received signal, and
$v\of{k}[n]$ is additive white Gaussian noise (AWGN).
Throughout, it will be assumed that the channel length $L$ obeys $L<N$.
The channel is ``sparse'' in the sense that \emph{exactly} $S$ of the $L$ 
channel taps $\{h\of{k}[l]\}_{l=0}^{L-1}$ are non-zero during each fading 
block $k$, where the indices of these non-zero taps, collected in 
the set $\mc{L}\of{k}$, can change with fading block $k$.
We will refer to this channel as ``strictly sparse'' when $S<L$, and as
``non-sparse'' when $S=L$. 
Furthermore, the channel is ``unknown'' in the sense that the transmitter 
and receiver \emph{do not} know the channel realizations, although they 
\emph{do} know the channel statistics, which are described as follows. 

Recalling that there are $M\defn\binom{L}{S}$ distinct $S$-element subsets 
of $\{0,\dots,L-1\}$, we write this collection of subsets as
$\{\mc{L}_i\}_{i=1}^{M}$.
We then assume that the channel support $\mc{L}\of{k}$ is drawn so that the 
event $\mc{L}\of{k}=\mc{L}_i$ occurs with prior probability $\lambda_i$, 
where $\mc{L}\of{k}$ is drawn 
independently of $\mc{L}\of{k'}$ for $k'\neq k$. 
We also assume that 
the vector $\vec{h}\of{k}\nz\in\Complex^{S}$ containing the non-zero 
taps $\{h\of{k}[l]:l\in\mc{L}\of{k}\}$ has the circular Gaussian
distribution\footnote{
For ease of presentation, we assume that all non-zero channel taps 
have equal variance.  All of our results except \lemref{optd} and \corref{optd} remain valid for any positive definite covariance matrix of 
$\vec{h}\of{k}\nz$, and both \lemref{optd} and \corref{optd} can be
straightforwardly extended to the general case.} 
$\vec{h}\of{k}\nz \sim \mc{CN}(\vec{0},S^{-1}\vec{I})$,
with $\vec{h}\of{k}\nz$ independent of $\vec{h}\of{k'}\nz$ for $k'\neq k$.
Finally, we assume that
$v\of{k}[n]\sim\mc{CN}(0,1)$ with $v\of{k}[n]$ independent of
$v\of{k'}[n']$ for $(k',n')\neq (k,n)$.
We impose the power constraint 
$\frac{1}{N}\sum_{n=0}^{N-1}\E\{|x\of{k}[n]|^2\}=1~\forall k$,
so that the signal-to-noise ratio (SNR) becomes $\rho$ in \eqref{ykn}.

Our channel model is motivated by the results of recent channel-sounding 
experiments (e.g., \cite{Cramer:TAP:02,Preisig:JAcSA:04,Molisch:TVT:05})
which suggest that, as the communication bandwidth increases, the 
channel taps $\{h\of{k}[n]\}_{n=0}^{L-1}$ become sparse in that 
the majority of them are ``below the noise floor'' \cite[p.\ 2]{Bajwa:PROC:10}.
The same behavior can be seen to manifest \cite{Schniter:JSTSP:11}
in channel taps sampled from IEEE~802.15.4a \cite{Molisch:802.15.4a} 
``ultra wideband'' propagation-path-based continuous-time impulse responses 
after square-root-raised-cosine pulse shaping.\footnote{
  Say that 
  $h\of{k}(t)=\sum_{q=1}^Q a_q e^{j\phi_q} \delta(t-\tau_q)$
  is a continuous-time impulse response based on $Q$ propagation paths.
  When the pulse shape $b_{\textsf{t}}(t)$ is used at the
  transmitter and $b_{\textsf{r}}(t)$ is used at the receiver, 
  and the baud interval is $T$, the channel taps become
  $h\of{k}[l] = (b_{\textsf{t}} \conv h\of{k} \conv b_{\textsf{r}})(lT)$,
  where $\conv$ denotes convolution.
  For a detailed derivation, see, e.g., \cite{Schniter:JSTSP:11}.} 
Clearly, the fact that we use \emph{exactly} zero-valued taps makes 
our channel model an approximation, albeit a standard one
(see, e.g., \cite[p.\ 5]{Bajwa:PROC:10}).
In fact, our channel model ignores many additional features\footnote{
    For example, in practice, 
    the active taps $\{h\of{k}[l]\}_{l\in\mc{L}\of{k}}$ 
    and additive noise might be non-Gaussian and/or correlated
    	within a fading block;
    the active taps, support, and noise might be statistically dependent 
    	and/or non-stationary across fading blocks; and
    the linear channel assumption might break down due to power-amplifier 
    	non-linearities.} 
of real-world channels in order to facilitate an 
information-theoretic analysis. 
In addition, it should be emphasized that we assume 
a channel with \emph{exactly} $S$ 
non-zero taps, as opposed to \emph{at most} $S$ non-zero taps,
and a decoder that knows the channel statistics perfectly
(including $S$, $L$, $\{\lambda_i\}_{i=1}^M$, and $\rho$).

 \emph{Notation}:
 Above and in the sequel, we use 
 lowercase boldface quantities to denote vectors, uppercase boldface 
 quantities to denote matrices, and we use
 $\vec{I}$ to denote the identity matrix.
 Also, we use
 $(\cdot)\tran$ to denote transpose,
 $(\cdot)^*$ conjugate, 
 $(\cdot)\herm$ conjugate transpose, 
 $(\cdot)^+$ pseudo-inverse, and 
 %
 $\Diag(\vec{b})$ the diagonal matrix created from vector $\vec{b}$.
 %
 Furthermore, 
 $\odot$ element-wise multiplication, 
 $\norm{\vec{x}} \defn \sqrt{\vec{x}\herm\vec{x}}$, and
 $\norm{\vec{x}}_{A} \defn \sqrt{\vec{x}\herm\vec{Ax}}$ 
	for Hermitian positive semi-definite $\vec{A}$.
 Throughout, ``$\log$'' denotes the base-$2$ logarithm.
 For random variables, we use
 $\E\{\cdot\}$ to denote expectation, 
 $\cov\{\vec{b}\}$ auto-covariance, 
 $\text{h}(\vec{a})$ entropy, and
 $\text{I}(\vec{a},\vec{b})$ the mutual information 
	between $\vec{a}$ and $\vec{b}$.
 %
 Finally,
 we write 
\textb{
 $\mc{CN}(\vec{x};\vec{\mu},\vec{\Sigma}) \defn
 (\pi^N \det(\vec{\Sigma}))^{-1}\exp(-\norm{\vec{x}-\vec{\mu}}_{\vec{\Sigma}^{-1}}^2)$
}
 for the circular Gaussian pdf with mean $\vec{\mu}\in\Complex^N$ and positive definite
 covariance matrix $\vec{\Sigma}$,
 and we write $\vec{x}\sim\mc{CN}(\vec{\mu},\vec{\Sigma})$ to indicate that
 random vector $\vec{x}$ has this pdf.
 In \tabref{def} we list commonly used quantities, along with their definitions.

\subsection{Preliminaries}				\label{sec:prelim}

Throughout the paper, we assume that 
the prefix samples 
$\{x\of{k}[-l]\}_{l=1}^{L-1}$ are chosen as a \emph{cyclic prefix} (CP), 
i.e., $x\of{k}[-l]=x\of{k}[N-l]$ for $l=1,\dots,L-1$.
In this case, we can write the $k^{th}$ block observations
$\vec{y}\of{k} \defn (y\of{k}[0],\dots,y\of{k}[N-1])\tran$ as
\begin{eqnarray}
 \vec{y}\of{k} 
 &=& \sqrt{\rho} \vec{X}\of{k}\vec{h}\of{k} + \vec{v}\of{k}, 
 							\label{eq:yXh}
\end{eqnarray}
where $\vec{v}\of{k} \defn (v\of{k}[0],\dots,v\of{k}[N-1])\tran$,
$\vec{h}\of{k}\defn (h\of{k}[0],\dots,h\of{k}[L-1],0,\dots,0)\tran 
\in \Complex^N$, and
$\vec{X}\of{k}\in\Complex^{N\times N}$ is the circulant matrix 
with first column
$\vec{x}\of{k} \defn (x\of{k}[0],\dots,x\of{k}[N-1])\tran$.
An equivalent model results\footnote{
  Model \eqref{yxh} follows directly from \eqref{yXh} using the fact that
  $\vec{X}\of{k} = \vec{F}\herm\Diag(\sqrt{N}\vec{Fx}\of{k})\vec{F}$.} 
from converting all signals into the frequency domain:
\begin{eqnarray}
 \vec{y}\of{k}\f
 &=& \sqrt{\rho} \Diag(\vec{x}\of{k}\f)\vec{h}\of{k}\f + \vec{v}\of{k}\f, 
 							\label{eq:yxh}
\end{eqnarray}
where 
$\vec{y}\of{k}\f \defn \vec{F}\vec{y}\of{k}$,
$\vec{x}\of{k}\f \defn \vec{F}\vec{x}\of{k}$,
$\vec{v}\of{k}\f \defn \vec{F}\vec{v}\of{k}$,
$\vec{h}\of{k}\f \defn \sqrt{N}\vec{F}\vec{h}\of{k}$,
and where
$\vec{F}$ denotes the $N$-dimensional unitary discrete Fourier transform 
(DFT) matrix. 
Noting that $\vec{v}\of{k}\f\sim\mc{CN}(\vec{0},\vec{I})$,
the model \eqref{yxh} establishes that, when viewed in the frequency domain,
the frequency-selective channel \eqref{yXh} reduces to a set of $N$
non-interfering scalar subchannels with average\footnote{
The average subchannel SNR of $\rho$ follows from the fact that 
$\frac{1}{N}\E\{\norm{\vec{h}\of{k}\f}^2\}=1$.}
subchannel SNR $\rho$.
Although the subchannels are non-interfering, the subchannel gains within the 
$k^{th}$ block (i.e., the elements of the vector $\vec{h}\of{k}\f$) 
are correlated in a way that depends on the channel support $\mc{L}\of{k}$, 
as will be detailed in the sequel.
For capacity analysis, we assume that the number of fading blocks $K$ 
is arbitrarily large, and we ignore overhead due to the prefix, 
consistent with \cite{Vikalo:TSP:04, Liang:TIT:04}.
Some implications of this choice are discussed below.

\subsection{Existing Results on Noncoherent Channel Capacity}

Much is known about the fundamental limits of reliable communication 
over the unknown \emph{non}-sparse channel in the high-SNR regime 
(i.e., $\rho\rightarrow\infty$). 
For example, assuming that communication occurs over an arbitrarily
large number of fading blocks $K$, the ergodic capacity 
$\mc{C}\nonsparse(\rho)$, in bits
per channel use, obeys \cite{Vikalo:TSP:04, Liang:TIT:04}
\begin{eqnarray}
 \lim_{\rho\rightarrow\infty} \frac{\mc{C}\nonsparse(\rho)}{\log \rho} 
 &=& 1-\frac{L}{N}.				\label{eq:Cnonsparse}
\end{eqnarray}
In other words, the ``multiplexing gain'' \cite{Tse:Book:05}
of the non-sparse channel
(i.e., the pre-log factor in its ergodic capacity expression) equals 
$1-\frac{L}{N}$.
Furthermore, it is possible to 
achieve this multiplexing gain using pilot aided transmission (PAT),
which uses $L$ signal-space dimensions of each fading block to 
transmit a known pilot signal and the remaining $N-L$ dimensions to 
transmit the data \cite{Vikalo:TSP:04, Liang:TIT:04}.
In the sequel, we use the term ``spectrally efficient'' to describe a
communication scheme whose achievable rate expression has a pre-log factor 
matching that of the channel's ergodic capacity expression (i.e., 
the channel's multiplexing gain).

\subsection{Our Contributions}			\label{sec:new}

In this paper, we study the fundamental limits of reliable communication
over the unknown \emph{sparse} channel \eqref{ykn} in the high-SNR regime.
First, we show that the ergodic capacity $\mc{C}\sparse(\rho)$ obeys
\begin{eqnarray}
 \lim_{\rho\rightarrow\infty} \frac{\mc{C}\sparse(\rho)}{\log \rho} 
 &=& 1-\frac{S}{N}				\label{eq:Csparse}
\end{eqnarray}
for any sparsity $S$ such that $1 \leq S \leq L < N$.
Comparing \eqref{Csparse} to \eqref{Cnonsparse}, it is interesting to
notice that the channel's multiplexing gain depends on the number of
non-zero taps $S$ and not the channel length $L$, even though
the locations of these taps
$\vec{\mc{L}} \defn (\mc{L}\of{1},\dots,\mc{L}\of{K})$ are unknown.
Second, we show that the sparse frequency-selective block-fading channel 
admits spectrally efficient PAT, just as its non-sparse variant does.
In other words, for an $S$-sparse channel, one can construct a PAT scheme 
that uses only $S$ pilots per fading block to attain an achievable rate 
that grows with SNR $\rho$ at the maximum possible rate, regardless of 
the channel length $L$. 
We establish this result constructively, by specifying a particular 
OFDM-based PAT scheme and a corresponding decoder, 
which---as we will see---can be interpreted as a 
\emph{joint channel-support/data decoder}. 
Because our decoder is computationally demanding 
(e.g., it requires the evaluation of up to $|\mc{L}| = M^K = 
\mc{O}(L^{SK})$ support hypotheses), 
we also consider a much simpler PAT decoder and
find that its $\epsilon$-achievable-rate has a pre-log factor 
of $1-\frac{S+1}{N}$, for any error-rate $\epsilon>0$.

In stating the above pre-log factors, we emphasize that the overhead due 
to the OFDM prefix has been ignored (for consistency with 
\cite{Vikalo:TSP:04, Liang:TIT:04}).
If, instead, the overhead was included, then the pre-log factor of the
non-sparse channel's ergodic capacity \eqref{Cnonsparse} would read 
as $\frac{N-L}{N+L-1}$, and that for the sparse channel \eqref{Csparse} 
would read as $\frac{N-S}{N+L-1}$.
Although the increase in pre-log factor resulting from channel sparsity, 
i.e., $\frac{L-S}{N+L-1}$, is not as pronounced as when the prefix is 
ignored, i.e., $\frac{L-S}{N}$, the two values are very similar when 
$N\gg L-1$, which is the typical case in practice.

\subsection{Relation to Compressed Channel Sensing}		\label{sec:ccs}

The problem of communicating over sparse channels has recently gained 
a significant amount of attention through the framework of 
\emph{compressed channel sensing} (CCS), as seen by the recent
overview article \cite{Bajwa:PROC:10} and the long list of papers 
cited therein.
In CCS, it is assumed that pilots are embedded during transmission, 
and that channel estimation is performed using pilot-only observations
(i.e., without the aid or interference from data).
CCS then exploits channel sparsity to reduce the number of pilots needed 
for accurate channel estimation, in the hopes of increasing spectral efficiency.
As an example, for the $N$-subcarrier OFDM scenario described by \eqref{yxh}, 
CCS results \cite{Bajwa:PROC:10} show that, 
when $P=\mc{O}(S_{\max}\ln^5 N)$ pilot subcarriers are chosen 
uniformly at random,
any \emph{deterministic} $L$-length channel $\vec{h}\of{k}$ 
with sparsity \emph{at most} $S_{\max}$
yields a CCS estimate $\hvec{h}\ccs\of{k}$ such that
\begin{eqnarray}
  \norm{\hvec{h}\ccs\of{k}-\vec{h}\of{k}}^2
  &\leq & C \frac{S_{\max} N \ln L}{\rho P} 
  	\emph{~~with high probability},
  \label{eq:ccs}
\end{eqnarray}
where $C$ is a constant.
The success probability in \eqref{ccs} grows with $L$ and $N$, but not with 
SNR $\rho$ (see \cite{Bajwa:PROC:10} for details).
Furthermore, in the special case that the observations are noise-free, 
it is known that exactly $2S$ data-free observations are both necessary 
and sufficient for perfect recovery \cite{Candes:TIT:06}. 

In comparing the CCS approach to the approach that we have taken,
we notice that the two are fundamentally different.
For example, CCS yields guarantees on the performance of 
channel estimation, but not on the rate of reliable communication.
Also, CCS attacks the channel estimation problem using a
``non-random parameter estimation'' framework, whereas we approach
channel estimation using a ``random parameter estimation'' framework,
since we consider ergodic capacity and achievable rate, and are thus
interested in average channel estimation performance.
A potential weakness to the CCS approach is that it uses only pilot 
observations for channel estimation, even though the data-dependent 
observations contain valuable information about the unknown channel;
our work (and related empirical results in 
\cite{Schniter:ASIL:10,Schniter:PHYCOM:11,Schniter:JSTSP:11}) 
suggests that significant gains can result from the use of 
\emph{joint} channel-estimation and data decoding. 
Strengths of CCS include the facts that
i) CCS focuses on reconstruction techniques that have polynomial complexity 
in $L$ and $S_{\max}$;
ii) CCS focuses on reconstruction techniques that do not need to know the
distributions of the signal and noise;
iii) CCS guarantees like \eqref{ccs}, which hold for \emph{any} sparsity
$S\leq S_{\max}$, can be further extended to cover the case of approximately 
sparse (i.e., ``compressible'') signals \cite[p.\ 5]{Bajwa:PROC:10}.

\section{Noncoherent Capacity}					\label{sec:cap}
In this section, we characterize the ergodic noncoherent capacity
of the sparse frequency-selective block-fading channel described 
in \secref{intro}.
We focus on the high-SNR regime, i.e., $\rho\rightarrow\infty$.


\begin{theorem}							\label{thm:cap}
The ergodic noncoherent capacity of the sparse frequency-selective 
block-fading channel, $C\sparse(\rho)$, in bits per channel use, obeys 
$\lim_{\rho \rightarrow \infty} {\frac{C\sparse(\rho)}{\log \rho}}
= 1-\frac{S}{N}$
for sparsity $S$ and block length $N$, whether or not the channel support 
realization $\vec{\mc{L}}$ is known apriori. 
\end{theorem}

\begin{IEEEproof}
Using the chain rule for mutual information \cite{Gallager:Book:68},
it follows straightforwardly that
\begin{eqnarray}
 \mi(\vec{y}\of{k}; \vec{x}\of{k}) 
 &=& \mi(\vec{y}\of{k};\mc{L}\of{k}) 
 	+ \mi(\vec{y}\of{k};\vec{x}\of{k}\giv\mc{L}\of{k}) 
	- \mi(\vec{y}\of{k};\mc{L}\of{k}\giv\vec{x}\of{k}). 	\label{eq:mie}
\end{eqnarray}
where $\mi(\vec{a};\vec{b})$ denotes the mutual information between
random vectors $\vec{a}$ and $\vec{b}$ and where 
$\mi(\vec{a};\vec{b}\giv\vec{c})$ denotes 
the conditional mutual information between $\vec{a}$ and $\vec{b}$
conditioned on $\vec{c}$.
Then, since $|\mc{L}\of{k}|=M$, we can bound the first
term in \eqref{mie} as follows:
\begin{eqnarray}
 \mi(\vec{y}\of{k};\mc{L}\of{k}) 
 &\leq& \en(\mc{L}\of{k})  
 ~\leq~ \log |\mc{L}\of{k}| 
 ~=~ \log \textstyle M,			 \label{eq:h0}
\end{eqnarray}
where $\en(\vec{a})$ denotes the entropy of $\vec{a}$.
Because $\mi(\vec{y}\of{k};\mc{L}\of{k}\giv\vec{x}\of{k}) \geq 0$, 
\eqref{mie}-\eqref{h0} yield the upper bound
$\mi(\vec{y}\of{k};\vec{x}\of{k}) 
 \leq \log M + \mi(\vec{y}\of{k};\vec{x}\of{k} \giv \mc{L}\of{k})$. 
Similarly, since $\mi(\vec{y}\of{k};\mc{L}\of{k})\geq 0$, equation
\eqref{mie} implies that
$\mi(\vec{y}\of{k};\vec{x}\of{k}) 
 \geq \mi(\vec{y}\of{k};\vec{x}\of{k} \giv \mc{L}\of{k}) - 
 \mi(\vec{y}\of{k};\mc{L}\of{k} \giv \vec{x}\of{k})$ 
and, since 
 $\mi(\vec{y}\of{k};\mc{L}\of{k}\giv\vec{x}\of{k}) 
  \leq \en(\mc{L}\of{k}\giv\vec{x}\of{k}) \leq \log M$, 
we also have that
$\mi(\vec{y}\of{k};\vec{x}\of{k}) 
 \geq  \mi(\vec{y}\of{k};\vec{x}\of{k} \giv \mc{L}\of{k}) - \log M$.
In summary, we have that
\begin{eqnarray}
  \mi(\vec{y}\of{k};\vec{x}\of{k})
  &=& \mi(\vec{y}\of{k};\vec{x}\of{k} \giv \mc{L}\of{k}) + \Delta
  	\text{~~for~ $\Delta \in \textstyle \big[-\log M,
		\log M\big]$} .		\label{eq:Delta}
\end{eqnarray}

Given knowledge of the support $\mc{L}\of{k}$,
the frequency-domain vector $\vec{h}\of{k}\f$ is zero-mean Gaussian with 
a rank-$S$ covariance matrix.
Thus, \cite[Theorem~1]{Liang:TIT:04} implies that 
$\mc{C}_{\mc{L}}(\rho)$, the pre-log factor of ergodic noncoherent capacity under knowledge
of the support $\vec{\mc{L}}$ equals $1-\frac{S}{N}$, i.e., 
$\lim_{\rho\rightarrow\infty} \frac{\mc{C}_{\mc{L}}(\rho)}{\log \rho} 
 = 1 - \frac{S}{N}$.
Since
\begin{eqnarray}
 \mc{C}_{\mc{L}}(\rho) 
 &=& \max_{p(\vec{x}\of{k}\f): \E\norm{\vec{x}\of{k}\f}^2 \leq N}
 \frac{1}{N}\mi(\vec{y}\of{k}\f;\vec{x}\of{k}\f\giv\mc{L}\of{k}), 
\end{eqnarray}
where 
$\mi(\vec{y}\of{k}\f;\vec{x}\of{k}\f \giv \mc{L}\of{k}) = 
\mi(\vec{y}\of{k};\vec{x}\of{k} \giv \mc{L}\of{k})$ 
and where, due to \eqref{Delta}, 
$\mi(\vec{y}\of{k};\vec{x}\of{k} \giv \mc{L}\of{k})$ differs from
$\mi(\vec{y}\of{k};\vec{x}\of{k})$ by a bounded $\rho$-invariant constant 
$\Delta$, the ergodic noncoherent capacity 
\begin{eqnarray}
 \mc{C}\sparse(\rho) 
 &=& \max_{p(\vec{x}\of{k}): \E\norm{\vec{x}\of{k}}^2 \leq N}
 \frac{1}{N}\mi(\vec{y}\of{k};\vec{x}\of{k}), 
\end{eqnarray}
must also obey $\lim_{\rho\rightarrow\infty} \frac{\mc{C}\sparse(\rho)}{\log \rho} 
= 1 - \frac{S}{N}$.
\end{IEEEproof} 

It is interesting to notice that the channel multiplexing gain equals
$1-\frac{S}{N}$ whether or not the support $\vec{\mc{L}}$ is apriori known.

\section{Pilot Aided Transmission and Decoding}		\label{sec:pat}

For the \emph{non-sparse} frequency-selective block-fading channel, it has 
been shown \cite{Vikalo:TSP:04}
that pilot aided transmission (PAT) is \emph{spectrally efficient} as defined 
in \secref{intro}, i.e., that it is possible to design a PAT scheme
for which the pre-log factor in its achievable rate expression coincides 
with the pre-log factor in the noncoherent ergodic capacity expression
(i.e., the channel multiplexing gain). 
The question remains as to whether PAT is spectrally efficient for 
\emph{sparse} channels as well.

Interestingly, \thmref{cap} showed that the multiplexing gain of the sparse
channel does not change with knowledge of the channel support $\vec{\mc{L}}$.
Realizing\footnote{
  The equivalence in pre-log factor between $S$-sparse channel with known support and
  a non-sparse length-$S$ channel follows directly from 
  \cite[Theorem~1]{Liang:TIT:04} and the fact that, in both cases,
  $\vec{h}\of{k}\f$ is zero-mean Gaussian with rank-$S$ covariance matrix.}
that an $S$-sparse channel with known support has
the same capacity characteristics as a non-sparse length-$S$ channel,
and recalling that PAT is spectrally efficient for non-sparse channels,
one might suspect that PAT is spectrally efficient for sparse channels.
As we shall see, this is indeed the case.
To prove this, we construct an appropriate PAT scheme and a corresponding
decoder, as detailed in the following subsections.

\subsection{PAT Definition}				\label{sec:patdefn}

For the transmission scheme outlined in \secref{prelim}, we consider a
PAT scheme in which the elements in the frequency-domain transmission 
vector $\vec{x}\f\of{k}\in\Complex^N$ 
can be partitioned into a pilot vector $\vec{x}\pl\in\Complex^P$,
created from $\{x\of{k}\f[n]: n\in\mc{N}\pl\}$, and a data 
vector $\vec{x}\da\of{k}\in\Complex^{N-P}$, created from 
$\{x\of{k}\f[n]: n\in\mc{N}\da\}$.
Here, we use $\mc{N}\pl \subset \{0,\dots,N-1\}$ to denote
the pilot subcarrier indices and 
$\mc{N}\da$ to denote the corresponding data subcarrier indices, where
$\mc{N}\da = \{0,\dots,N-1\}\setminus \mc{N}\pl$.
Notice that exactly $P$ signal-space dimensions (per fading block)
have been allocated to pilots, i.e., $|\mc{N}\pl|=P$. 
For simplicity, we assume that the pilot locations $\mc{N}\pl$ and pilot 
values $\vec{x}\pl$ do not change with the fading block $k$, and that
the pilot values are constant modulus, i.e., $|x\pl[n]|=1$.
By definition, the pilot quantities $\vec{x}\pl$ and $\mc{N}\pl$ are known 
apriori to the decoder.

In the parallel subchannel model \eqref{yxh}, we partition
both $\vec{y}\f\of{k}\in\Complex^N$ and $\vec{v}\f\of{k}\in\Complex^N$
in the same way as we did $\vec{x}\f\of{k}\in\Complex^N$, yielding 
\begin{eqnarray}
  \vec{y}\pl\of{k}
  &=& \sqrt{\rho} 
  	\Diag(\vec{x}\pl) \vec{J}\pl 
	\vec{h}\of{k}\f + \vec{v}\of{k}\pl 		\label{eq:ypl1}\\
  \vec{y}\da\of{k}
  &=& \sqrt{\rho} 
  	\Diag(\vec{x}\da\of{k}) \vec{J}\da 
	\vec{h}\of{k}\f + \vec{v}\of{k}\da ,		\label{eq:yda1}
\end{eqnarray}
where $\vec{J}\pl$ is a selection matrix constructed from rows
$\mc{N}\pl$ of the $N\times N$ identity matrix, and 
$\vec{J}\da$ is constructed similarly from rows
$\mc{N}\da$ of the identity matrix.
Another way to write $\vec{y}\pl\of{k}$ and $\vec{y}\da\of{k}$, which
will be useful in the sequel, is
\begin{eqnarray}
  \vec{y}\pl\of{k}
  &=& \sqrt{\rho N} \Diag(\vec{x}\pl) \vec{F}\of{k}\plt 
	\vec{h}\of{k}\nz + \vec{v}\of{k}\pl 		\label{eq:ypl2}\\
  \vec{y}\da\of{k}
  &=& \sqrt{\rho N} \Diag(\vec{x}\da\of{k}) \vec{F}\of{k}\dat 
	\vec{h}\of{k}\nz + \vec{v}\of{k}\da ,		\label{eq:yda2}
\end{eqnarray}
where
$\vec{h}\nz\of{k}\in\Complex^S$ is formed from the non-zero elements 
of $\vec{h}\of{k}$, 
$\vec{F}\of{k}\plt$ is formed from rows $\mc{N}\pl$ and columns 
$\mc{L}\of{k}$ of the DFT matrix $\vec{F}$, and 
$\vec{F}\of{k}\dat$ is formed from rows $\mc{N}\da$ and columns 
$\mc{L}\of{k}$ of $\vec{F}$.
Notice that, because $\mc{L}\of{k}$ is not apriori known to the decoder, 
neither are $\vec{F}\of{k}\plt$ or $\vec{F}\of{k}\dat$.


To achieve an arbitrarily small probability of decoding error, 
we construct codewords that span $K$ blocks, where $K$ is arbitrarily large. 
Thus, using $\code\subset\Complex^{K (N-P)}$ to denote our codebook,
we partition each codeword $\vec{x}\da\in\code$ into $K$ data vectors, i.e.,
$\vec{x}\da=[\vec{x}\oft{1}\da,\dots,\vec{x}\oft{K}\da]\tran$, for use in 
our PAT scheme.
The codewords $\vec{x}\da$ are generated independently from a Gaussian
distribution such that the $\vec{x}\of{k}\da$ has positive definite
covariance matrix $\vec{R}\da$ for all $k$, and such that $\vec{x}\of{k}\da$ 
is independent of $\vec{x}\of{k'}\da$ for $k \neq k'$. 
Denoting the number of codewords in the codebook by $|\code|$, 
the average data rate is given by $\mc{R} = \frac{1}{KN} \log |\code|$.  

\begin{table}[p]
\centering
\begin{tabular}{|l|l|}\hline
  $\vec{y}\of{k},\vec{y}\f\of{k}$ & observation vector in time domain, in frequency domain \\
  $\vec{h}\of{k},\vec{h}\f\of{k}$ & channel vector in time domain, in frequency domain \\
  $\vec{x}\of{k},\vec{x}\f\of{k}$ & data vector in time domain, in frequency domain \\
  $\vec{v}\of{k},\vec{v}\f\of{k}$ & AWGN vector in time domain, in frequency domain \\
  $\vec{y}\pl\of{k},\vec{y}\da\of{k}$ & pilot, data portions of in frequency-domain observation vector\\
  $\vec{x}\pl\of{k},\vec{x}\da\of{k}$ & pilot, data portions of in frequency-domain data vector\\
  $\vec{v}\pl\of{k},\vec{v}\da\of{k}$ & pilot, data portions of in frequency-domain noise vector\\
  $\mc{N}\pl,\,\mc{N}\da$ & pilot, data subcarrier index sets \\
  $\vec{h}\nz\of{k}$ & non-zero portion of time-domain channel vector\\
  $\mc{L}\of{k}$ & set of channel-support indices for $k^{th}$ block\\
  $\mc{L}_i$ & set of channel-support indices for $i^{th}$ hypothesis\\
  $\vec{F}\of{k}\plt$ & unitary DFT matrix restricted to true columns $\mc{L}\of{k}$ and rows $\mc{N}\pl$\\
  $\vec{F}_i$ & unitary DFT matrix restricted to columns $\mc{L}_i$\\
  $\vec{F}\pli$ & unitary DFT matrix restricted to pilot rows $\mc{N}\pl$ and columns $\mc{L}_i$\\
  $\vec{F}\di$ & unitary DFT matrix restricted to data rows $\mc{N}\da$ and columns $\mc{L}_i$\\
  $\hvec{h}\fpi\of{k}, \tvec{h}\fpi\of{k}$ & $\mc{L}_i$-conditional pilot-based MMSE estimate of $\vec{h}\f\of{k}$, associated error\\
  $\hvec{h}\nzpi\of{k}, \tvec{h}\nzpi\of{k}$ & $\mc{L}_i$-conditional pilot-based MMSE estimate of $\vec{h}\nz\of{k}$, associated error\\
  $\vec{e}\di\of{k}$ & $\mc{L}_i$-conditional effective noise on $\vec{y}\da\of{k}$ for WMD decoding \\
  $\vec{z}\pl\of{k}$ & normalized pilot observations used for PASE\\
  $\vec{\nu}\pl\of{k}$ & normalized AWGN on $\vec{z}\pl\of{k}$ used for PASE\\
  $\vec{e}\pli\of{k}$ & $\mc{L}_i$-conditional projection error vector used for PASE\\
\hline
\end{tabular}
\caption{Review of commonly used variables, where  
$(\cdot)\of{k}$ denotes dependence on $k^{th}$ fading block.}
\label{tab:def}
\end{table}

\subsection{Optimal Decoding for PAT}			\label{sec:optd}

The reader may naturally wonder: what is the \emph{optimal} decoder for 
the above PAT scheme in the case of the sparse channel described in 
\secref{intro}, and how does it compare to optimal decoding in the 
non-sparse case?
To answer these questions, we detail the optimal decoder for the 
sparse and non-sparse cases below.
In the sequel, we use
$\vec{F}_i\in\Complex^{N\times S}$ to denote the matrix formed 
from columns $\mc{L}_i$ of the DFT matrix $\vec{F}$,
we use $\vec{F}\pli\in\Complex^{P\times S}$ 
to denote the matrix formed from rows $\mc{N}\pl$ of $\vec{F}_i$, and
we use $\vec{F}\di\in\Complex^{(N-P)\times S}$ 
to denote the matrix formed from rows $\mc{N}\da$ of $\vec{F}_i$.

\begin{lemma}						\label{lem:optd}
The maximum likelihood decoder for PAT over the $S$-sparse $L$-length
frequency-selective $N$-block-fading channel takes the form
\begin{eqnarray}
  \hvec{x}\da\ml
  &=& \argmax_{\vec{x}\da\in\code} \prod_{k=1}^K \sum_{i=1}^M 
  	\hat{\lambda}\of{k}\pli
  	\det\Big(\rho N \vec{F}\di\herm 
		\Diag(\vec{x}\da\of{k}\odot\vec{x}\da\ofc{k})
		\vec{F}\di + \vec{\Sigma}\nzpi^{-1} \Big)^{-1}
  \nonumber\\&&\mbox{}
  \exp\Big(-\big\|\vec{y}\of{k}\da-\sqrt{\rho N}\Diag(\vec{x}\of{k}\da)
  	\vec{F}\di \hvec{h}\of{k}\nzi(\vec{x}\of{k}\da) \big\|^2  
	- \big\|\hvec{h}\of{k}\nzi(\vec{x}\of{k}\da) - \hvec{h}\of{k}\nzpi
	\big\|^2_{\vec{\Sigma}^{-1}\nzpi} \Big)	\quad
							\label{eq:optd}
\end{eqnarray}
where $\hat{\lambda}\of{k}\pli \defn 
\Pr\{\mc{L}\of{k}=\mc{L}_i \giv \vec{y}\of{k}\pl, \vec{x}\pl\}$
is the pilot-aided channel-support posterior, where 
$\hvec{h}\of{k}\nzpi$ is the $\mc{L}_i$-conditional pilot-aided MMSE estimate
of $\vec{h}\of{k}\nz$ and $\vec{\Sigma}\nzpi$ is its error covariance,
which take the form
\begin{eqnarray}
  \hvec{h}\of{k}\nzpi
  &=& \textstyle \sqrt{\frac{\rho}{N}}\vec{F}\pli\herm
  	\big( \rho\vec{F}\pli\vec{F}\pli\herm 
	+ \frac{S}{N}\vec{I}\big)^{-1} \Diag(\vec{x}\pl^*)
	\vec{y}\of{k}\pl ,				\label{eq:hnzpi} \\
  \vec{\Sigma}\nzpi
  &=& \textstyle \frac{1}{S} \big( \vec{I} 
  	- \vec{F}\pli\herm
	\big( \vec{F}\pli\vec{F}\pli\herm 
        	+ \frac{S}{\rho N}\vec{I}\big)^{-1}
	\vec{F}\pli \big) ,  				\label{eq:Signzpi}
\end{eqnarray}
and where $\hvec{h}\of{k}\nzi(\vec{x}\of{k}\da)$
denotes the MMSE estimate of $\vec{h}\of{k}\nz$ conditioned on the 
data hypothesis $\vec{x}\of{k}\da$ and based on the pilot-aided channel
statistics \eqref{hnzpi}-\eqref{Signzpi}, i.e.,
\begin{eqnarray}
  \hvec{h}\of{k}\nzi(\vec{x}\of{k}\da)
  &=& \hvec{h}\of{k}\nzpi + \sqrt{\rho N}\vec{\Sigma}\nzpi 
  	\vec{F}\di\herm\Diag(\vec{x}\ofc{k}\da)
	\big(\rho N\Diag(\vec{x}\of{k}\da)\vec{F}\di
		\vec{\Sigma}\nzpi 
		\vec{F}\di\herm\Diag(\vec{x}\ofc{k}\da)
	+ \vec{I} \big)^{-1}
	\nonumber\\&&\mbox{}\times
	\big(\vec{y}\of{k}\da 
		- \sqrt{\rho N}\Diag(\vec{x}\of{k}\da)\vec{F}\di 
		\hvec{h}\of{k}\nzpi \big) .		\label{eq:hnzdi}
\end{eqnarray}
\end{lemma}
\begin{IEEEproof}
See \appref{optd}.
\end{IEEEproof}
Paraphrasing \lemref{optd}, the optimal decoder \eqref{optd} for 
sparse-channel PAT first uses pilots to compute support posteriors 
$\{\hat{\lambda}\of{k}\pli\}_{i=1}^M$
and support-conditional channel posteriors\footnote{
  Note that $\{\vec{\Sigma}\nzpi\}_{i=1}^M$ can be precomputed 
  since they do not depend on the observations.}
$\{\hvec{h}\of{k}\nzpi\}_{i=1}^M$ for each fading block $k$.
Then, it averages over the $M$ support hypotheses to obtain a joint 
data-channel decoding metric for each fading block $k$.
Finally, it searches for the codeword that maximizes the product of the 
decoding metrics (over all fading blocks $k$).
We note that optimal decoding is an example of Bayes model averaging
\cite{Clyde:SS:04} and differs markedly from the decoding approach
implied in the compressed channel sensing (CCS) framework \cite{Bajwa:PROC:10},
which aims to compute a \emph{single} sparse channel estimate 
$\{\hvec{h}\of{k}\nzpi, \mc{L}\of{k}=\mc{L}_i\}$ for later use in data 
decoding.
We also note that ML decoding complexity is\footnote{
  The term after the sum in \eqref{optd} must be computed
  for every triple $(i,k,\vec{x}\of{k}\da)$, where the complexity of
  each computation is $\mc{O}(N^3)$ due to the matrix inversion in 
  \eqref{hnzdi}.
  } 
$\mc{O}(|\code| M K N^3)$. 

For illustrative purposes, we compare the optimal decoder for a 
sparse channel (as specified in \lemref{optd} above) to the optimal 
decoder for a \emph{non}-sparse channel, as detailed below in \corref{optd}.

\begin{corollary}					\label{cor:optd}
The maximum likelihood decoder for PAT over the non-sparse $L$-length 
frequency-selective $N$-block-fading channel takes the form
\newcommand{\done}{_{\text{\sf d}}}
\newcommand{\plone}{_{\text{\sf p}}}
\begin{eqnarray}
  \hvec{x}\da\ml
  &=& \argmin_{\vec{x}\da\in\code} \sum_{k=1}^K 
  \Big( \ln\det\big(\rho N \vec{F}\done\herm
  	\diag(\vec{x}\da\of{k}\odot\vec{x}\da\ofc{k}) \vec{F}\done
	+ \vec{\Sigma}\nzp^{-1} \big)
	\nonumber\\&&\mbox{}
   	+ \big\|\vec{y}\of{k}\da-\sqrt{\rho N}\Diag(\vec{x}\of{k}\da)
  	\vec{F}\done \hvec{h}\of{k}\nz(\vec{x}\of{k}\da) \big\|^2  
	+ \big\|\hvec{h}\of{k}\nz(\vec{x}\of{k}\da) - \hvec{h}\of{k}\nzp
	\big\|^2_{\vec{\Sigma}^{-1}\nzp} \Big)	,
							\label{eq:optd_non}
\end{eqnarray}
where $\hvec{h}\of{k}\nzp$ is the pilot-aided MMSE estimate
of $\vec{h}\of{k}\nz$ and $\vec{\Sigma}\nzp$ is its error covariance,
which take the form
\begin{eqnarray}
  \hvec{h}\of{k}\nzp
  &\defn& \textstyle \sqrt{\frac{\rho}{N}}\vec{F}\plone\herm
  	\big( \rho\vec{F}\plone\vec{F}\plone\herm 
	+ \frac{L}{N}\vec{I}\big)^{-1} \Diag(\vec{x}\pl^*)
	\vec{y}\of{k}\pl ,				\label{eq:hnzp1}\\
  \vec{\Sigma}\nzp
  &\defn& \textstyle \frac{1}{L}\big(\vec{I} - \vec{F}\plone\herm
	\big( \vec{F}\plone\vec{F}\plone\herm 
        	+ \frac{L}{\rho N}\vec{I}\big)^{-1}
	\vec{F}\plone \big) ,  				\label{eq:Signzp1}
\end{eqnarray}
and where $\hvec{h}\of{k}\nz(\vec{x}\of{k}\da)$
denotes the MMSE estimate of $\vec{h}\of{k}\nz$ conditioned on the 
data hypothesis $\vec{x}\of{k}\da$ and based on the pilot-aided channel
statistics \eqref{hnzp1}-\eqref{Signzp1}, i.e.,
\begin{eqnarray}
  \hvec{h}\of{k}\nz(\vec{x}\of{k}\da)
  &=& \hvec{h}\of{k}\nzp + \sqrt{\rho N}\vec{\Sigma}\nzp 
  	\vec{F}\done\herm\Diag(\vec{x}\ofc{k}\da)
	\big(\rho N\Diag(\vec{x}\of{k}\da)\vec{F}\done
		\vec{\Sigma}\nzp 
		\vec{F}\done\herm\Diag(\vec{x}\ofc{k}\da)
	+ \vec{I} \big)^{-1}
	\nonumber\\&&\mbox{}\times
	\big(\vec{y}\of{k}\da 
		- \sqrt{\rho N}\Diag(\vec{x}\of{k}\da)\vec{F}\done 
		\hvec{h}\of{k}\nzp \big) .
\end{eqnarray}
\end{corollary}
To paraphrase \corref{optd}, the optimal decoder \eqref{optd} for 
\emph{non}-sparse-channel PAT computes a single pilot-aided MMSE channel 
estimate $\hvec{h}\of{k}\nzp$,
which is then used to construct a joint data-channel 
decoding metric, for each fading block $k$.
Finally, it searches for the codeword \textb{that minimizes} the sum of the 
decoding metrics (over $k$).
It can be seen that optimal decoding in the sparse case differs from
that in the non-sparse cases by the need to compute, at each fading
block $k$, the support posteriors 
$\{\hat{\lambda}\of{k}\pli\}_{i=1}^M$ and the corresponding
support-conditional tap estimates $\{\hvec{h}\of{k}\nzpi\}_{i=1}^M$
and then average the decoding metrics over the $M$ support hypotheses.

\subsection{Decoupled Decoding of PAT}         		\label{sec:decoupled}

For both sparse and non-sparse channels, the optimal decoder of PAT, 
as detailed in \secref{optd}, takes the form of a joint-{channel/data} decoder.
In practice, for reasons of simplicity,
decoding is often \emph{decoupled} into two stages: 
i) pilot-aided channel estimation and
ii) coherent data-decoding based on the channel estimate.
We now detail a decoupled decoder for the sparse channel 
of \secref{intro} and the PAT scheme of \secref{pat} 
that, while suboptimal, performs well enough to yield spectrally efficient
communication \emph{when provided with the correct value of the
channel support $\vec{\mc{L}}$}.
In the sequel, we will refer to the case of known 
$\vec{\mc{L}}$ as the \emph{support-genie} case.
Later, in Sections~\ref{sec:dasd}~and~\ref{sec:pasd}, we will propose 
schemes to reliably infer the support $\vec{\mc{L}}$.

For our decoupled decoder, pilot-aided channel estimation is accomplished 
in a \emph{support-hypothesized} manner.
More precisely, we compute---at each fading block $k$---the pilot-aided 
MMSE estimate $\hvec{h}\of{k}\fpik$ of the non-zero taps 
$\vec{h}\of{k}\f$ under channel-support hypothesis $\mc{L}\of{k}=\mc{L}\ik$.
To do this, we set 
$\hvec{h}\of{k}\fpik=\sqrt{N}\vec{F}\ik\hvec{h}\of{k}\nzpik$ 
for the $\hvec{h}\of{k}\nzpik$ specified by \eqref{hnzpi}. 
Note that $\hvec{h}\of{k}\fpik$ is a linear estimate due to the fact
that $\vec{h}\of{k}\f$ becomes Gaussian when conditioned on a particular 
support. 
In contrast, the (support-unconditional) pilot-aided MMSE estimate of 
$\vec{h}\of{k}\f$ is in general non-linear.
%
%
The support-hypothesized channel 
estimates $\{\hvec{h}\of{k}\fpik\}_{k=1}^K$ and their
covariances $\{\vec{\Sigma}\fpik\}_{k=1}^K$ 
are then used in coherent data decoding.
(Note that $\vec{\Sigma}\fpik = N\vec{F}\ik\vec{\Sigma}\nzpik\vec{F}\ik\herm$,
where $\vec{\Sigma}\nzpik$ is given by \eqref{Signzpi}).
\textb{
For coherent data decoding, we employ the weighted minimum-distance 
(WMD) decoder, defined \cite{Weingarten:TIT:04} as
\begin{eqnarray}
 \hvec{x}\dii\wmd 
 &=& \argmin_{\vec{x}\da \in \code} \sum_{k=1}^{K} 
 \big\|\vec{Q}\of{k}\ik 
 	\big(\vec{y}\of{k}\da - \sqrt{\rho} \Diag(\vec{x}\of{k}\da)
 	\vec{J}\da\hvec{h}\of{k}\fpik \big)\big\|^2, 	\label{eq:deco}
\end{eqnarray}
where $\vec{Q}\of{k}\ik$ is a weighting matrix and $\vec{i}=(i_1,\dots,i_K)$. 
Writing the observation as 
\begin{eqnarray}
 \vec{y}\of{k}\da
 &=& \sqrt{\rho}\Diag(\vec{x}\of{k}\da)\vec{J}\da \hvec{h}\of{k}\fpik +
 	\underbrace{ 
	\sqrt{\rho}\Diag(\vec{x}\of{k}\da)\vec{J}\da \tvec{h}\of{k}\fpik
	+ \vec{v}\of{k}\da 
	}_{\displaystyle \defn \vec{e}\of{k}\dik},
							\label{eq:edi}
\end{eqnarray}
the standard \cite{Weingarten:TIT:04} choice for $\vec{Q}\of{k}\ik$ is a 
whitening matrix for the ``effective noise'' $\vec{e}\of{k}\dik$.
We note that the covariance 
$\vec{C}\dik\defn\cov\{\vec{e}\of{k}\dik\}$ 
(and thus $\vec{Q}\of{k}\ik$) depends on $\vec{\Sigma}\fpik$, $\vec{R}\da$, 
and $\rho$.
}

%

For the achievable rate of the decoupled-decoder PAT system to grow
logarithmically with $\rho$, the effective noise $\vec{e}\of{k}\dik$
must satisfy certain properties. 
Towards this aim, we establish that, with $P \geq S$ pilot tones, 
the support hypothesized channel estimation error variance decays 
at the rate of $\frac{1}{\rho}$ as $\rho\rightarrow\infty$,
if and only if the support hypothesis is correct.

\begin{lemma}						\label{lem:lmmse}
Say that $N$ is prime. 
Then, for any pilot pattern $\mc{N}\pl$ such that $P\geq S$,
there exists a constant $C$ such that the channel estimation error 
obeys $\E\{\norm{\tvec{h}\of{k}\fpi}^2\} \leq C\rho^{-1}$
for all $\rho>0$ if and only if $\mc{L}_i = \mc{L}\of{k}\true$, i.e., $\mc{L}_i$ is
the true channel-support of $k^{th}$ block.
\end{lemma}
\begin{IEEEproof}
We begin by recalling that, under support hypothesis $\mc{L}\of{k}=\mc{L}_i$, 
the frequency-domain channel coefficients $\vec{h}\of{k}\f$ are related 
to the non-zero channel taps $\vec{h}\of{k}\nz$
via $\vec{h}\of{k}\f = \sqrt{N}\vec{F}_i \vec{h}\of{k}\nz$, where 
$\vec{F}_i$ contains columns $\mc{L}_i$ of the unitary DFT matrix $\vec{F}$.
Thus, $\hvec{h}\of{k}\fpi$, the $\mc{L}_i$-conditional pilot-aided MMSE 
estimate of $\vec{h}\of{k}\f$ is related to $\hvec{h}\of{k}\nzpi$, the 
$\mc{L}_i$-conditional MMSE pilot-aided estimate of $\vec{h}\of{k}\nz$, via 
$\hvec{h}\of{k}\fpi = \sqrt{N} \vec{F}_i \hvec{h}\of{k}\nzpi$.
Because the columns of $\vec{F}_i$ are orthonormal, the estimation error obeys
\begin{eqnarray}
  \norm{\tvec{h}\of{k}\fpi}^2
  ~=~ \norm{\vec{h}\of{k}\f-\hvec{h}\of{k}\fpi}^2
  &=& N \norm{\vec{h}\of{k}\nz-\hvec{h}\of{k}\nzpi}^2 
  ~=~ N \norm{\tvec{h}\of{k}\nzpi}^2 .
\end{eqnarray}
%
Plugging \eqref{ypl2} into \eqref{hnzpi}, the estimation error 
$\tvec{h}\of{k}\nzpi \defn \vec{h}\of{k}\nz - \hvec{h}\of{k}\nzpi$ becomes
\begin{eqnarray}
 \tvec{h}\of{k}\nzpi
 &=& \textstyle 
 	\Big(\vec{I} - \vec{F}\pli\herm
 		\big(\vec{F}\pli\vec{F}\pli\herm
 		+ \frac{S}{\rho N}\vec{I} \big)^{-1} 
		\vec{F}\of{k}\plt \Big) \vec{h}\of{k}\nz 
 	\nonumber\\&&\mbox{}\textstyle
 	- \frac{1}{\sqrt{\rho N}}\vec{F}\pli\herm
 		\big(\vec{F}\pli\vec{F}\pli\herm
 		+ \frac{S}{\rho N}\vec{I}\big)^{-1}
 		\Diag(\vec{x}\pl^*)\vec{v}\of{k}\pl .	
\end{eqnarray}
Then, since $\vec{h}\of{k}\nz$ is independent of $\vec{v}\of{k}\pl$, 
\begin{eqnarray}
 \E\{\norm{\tvec{h}\of{k}\nzpi}^2\}
 &=& \textstyle \frac{1}{S}\tr\Big\{
 	\Big(\vec{I} - \vec{F}\pli\herm
 		\big(\vec{F}\pli\vec{F}\pli\herm
 		+ \frac{S}{\rho N}\vec{I} \big)^{-1} 
		\vec{F}\of{k}\plt \Big)
 	\nonumber\\&&\mbox{}\textstyle\times
 	\Big(\vec{I} - \vec{F}\pli\herm
 		\big(\vec{F}\pli\vec{F}\pli\herm
 		+ \frac{S}{\rho N}\vec{I} \big)^{-1} 
		\vec{F}\of{k}\plt \Big)\herm \Big\}
 	\nonumber\\&&\mbox{}\textstyle
 	+ \frac{1}{\rho N} \tr\Big\{ 
		\vec{F}\pli\herm
 		\big(\vec{F}\pli\vec{F}\pli\herm
 		+ \frac{S}{\rho N}\vec{I}\big)^{-2}
		\vec{F}\pli \Big\} .
\end{eqnarray}

We now make a few observations about $\vec{F}\pli$ and $\vec{F}\of{k}\plt$.
When $N$ is prime, the Chebotarev theorem 
\cite{Stevenhagen:MI:96,Tao:MRL:05}
guarantees that any square submatrix of the $N$-DFT matrix $\vec{F}$ will be 
full rank. 
Hence, any tall submatrix of $\vec{F}$ will also be full rank.
Then, because $P\geq S$, it follows that $\vec{F}\pli\in\Complex^{P\times S}$ 
will be full rank for all $i$, as will $\vec{F}\of{k}\plt$.
Furthermore, when $\mc{L}_i \neq \mc{L}\of{k}\true$, it follows that 
$\vec{F}\pli \neq \vec{F}\of{k}\plt$.

To proceed, we use the singular value decomposition 
$\vec{F}\pli = \vec{U}_i\vec{\Sigma}_i \vec{V}_i\herm$,
where $\vec{\Sigma}_i\in\Complex^{P\times S}$ is a full-rank diagonal matrix
and where $\vec{U}_i$ and $\vec{V}_i$ are both unitary.
Then
\begin{eqnarray}
 \textstyle
 \vec{F}\pli\herm \big(\vec{F}\pli\vec{F}\pli\herm 
 + \frac{S}{\rho N}\vec{I}\big)^{-1} 
 &=& \vec{V}_i \underbrace{ \textstyle
 	\vec{\Sigma}_i\herm \big(\vec{\Sigma}_i\vec{\Sigma}_i\herm
		+ \frac{S}{\rho N} \vec{I} \big)^{-1}  
	}_{\displaystyle \defn \vec{D}_i\herm}
	\vec{U}_i\herm ,
\end{eqnarray}
where $\vec{D}_i \in \Complex^{P\times S}$ is full-rank diagonal with
non-zero elements 
$\{\frac{\sigma_{i,l}}{\sigma_{i,l}^2 + S/(\rho N)}\}_{l=1}^S$, using
$\sigma_{i,l}$ to denote the $l^{th}$ singular value in $\vec{\Sigma}_i$.

In the case that $\mc{L}_i=\mc{L}\of{k}\true$, we have 
$\vec{F}\of{k}\plt = \vec{F}\pli$, and so 
\begin{eqnarray}
 \E\{\norm{\tvec{h}\of{k}\nzpi}^2\}
 &=& \textstyle \frac{1}{S} \tr\big\{ 
 	(\vec{I}-\vec{V}_i\vec{D}_i\herm \vec{\Sigma}_i\vec{V}_i\herm)
 	(\vec{I}-\vec{V}_i\vec{\Sigma}_i\herm \vec{D}_i\vec{V}_i\herm)
 	\big\}
 	\nonumber\\&&\mbox{}\textstyle
	+ \frac{1}{\rho N} \tr\big\{
	\vec{V}_i\vec{D}_i\herm \vec{D}_i\vec{V}_i\herm
	\big\} \\
 &=& \textstyle \frac{1}{S} \tr\big\{ 
 	(\vec{I}-\vec{D}_i\herm \vec{\Sigma}_i)
 	(\vec{I}-\vec{\Sigma}_i\herm \vec{D}_i)
 	\big\}
	+ \frac{1}{\rho N} \tr\big\{
	\vec{D}_i\herm \vec{D}_i
	\big\} \\
 &=& \frac{1}{S} \sum_{l=1}^S 
 	\Big( 1 - \frac{\sigma_{i,l}^2}{\sigma_{i,l}^2 + S/(\rho N)} \Big)^2
	+ \frac{1}{\rho N} \sum_{l=1}^S
	\frac{\sigma_{i,l}^2}{(\sigma_{i,l}^2 + S/(\rho N))^2} \\
 &=& \sum_{l=1}^S \frac{1}{N \sigma_{i,l}^2 \rho + S} \\
 &\leq& \rho^{-1} \sum_{l=1}^S \frac{1}{N \sigma_{i,l}^2} .
\end{eqnarray}
Thus, we have the upper bound 
$\E\{\norm{\tvec{h}\of{k}\fpi}^2\} = N \E\{\norm{\tvec{h}\of{k}\nzpi}^2\}
\leq C \rho^{-1}$ with $C = \sum_{l=1}^S \sigma_{i,l}^{-2}$.

For the case $\mc{L}_i \neq \mc{L}\of{k}\true$, we have 
$\vec{F}\of{k}\plt \neq \vec{F}\pli$, and so we can use 
the previously defined SVD quantities to write 
$\vec{F}\of{k}\plt = \vec{U}_i(\vec{\Sigma}_i+\vec{\Delta}_i)\vec{V}_i^H$,
where $\vec{\Delta}_i \in \Complex^{P\times S}$ is some non-zero matrix.
It then follows that
\begin{eqnarray}
 \E\{\norm{\tvec{h}\of{k}\nzpi}^2\}
 &=& \textstyle \frac{1}{S} \tr\big\{ 
 	\big(\vec{I}-\vec{V}_i\vec{D}_i\herm 
	(\vec{\Sigma}_i+\vec{\Delta}_i)\vec{V}_i\herm\big)
 	\big(\vec{I}-\vec{V}_i
	(\vec{\Sigma}_i+\vec{\Delta}_i)\herm \vec{D}_i\vec{V}_i\herm\big)
 	\big\}
 	\nonumber\\&&\mbox{}\textstyle
	+ \frac{1}{\rho N} \tr\big\{
	\vec{V}_i\vec{D}_i\herm \vec{D}_i\vec{V}_i\herm
	\big\} \\
 &=& \textstyle \frac{1}{S} \tr\big\{ 
 	(\vec{I} - \vec{D}_i\herm\vec{\Sigma}_i - \vec{D}_i\herm\vec{\Delta}_i)
 	(\vec{I} - \vec{\Sigma}_i\herm\vec{D}_i - \vec{\Delta}_i\herm\vec{D}_i)
 	\big\}
 	\nonumber\\&&\mbox{}\textstyle
	+ \frac{1}{\rho N} \tr\big\{
	\vec{D}_i\herm \vec{D}_i
	\big\} \\
 &=& \E\{\norm{\tvec{h}\of{k}\nzt}^2\}
 	- \textstyle \frac{1}{S} \tr\big\{
	(\vec{I} - \vec{D}_i\herm\vec{\Sigma}_i)\vec{\Delta}_i\herm\vec{D}_i
	+\vec{D}_i\herm\vec{\Delta}_i(\vec{I} - \vec{\Sigma}_i\herm\vec{D}_i)
	\big\}
 	\nonumber\\&&\mbox{}\textstyle
 	+ \textstyle \frac{1}{S} \tr\big\{
	\vec{D}_i\herm\vec{\Delta}_i \vec{\Delta}_i\herm\vec{D}_i
	\big\}						\label{eq:Ethnzi}
\end{eqnarray}
As established above, $\E\{\norm{\tvec{h}\of{k}\nzt}^2\} \rightarrow 0$
as $\rho\rightarrow \infty$.
Since $\vec{I} - \vec{D}_i\herm\vec{\Sigma}_i$ is diagonal with elements
$\{\frac{1}{1+\rho N\sigma_{i,l}^2}\}_{l=1}^S$, the second term in 
\eqref{Ethnzi} also vanishes as $\rho\rightarrow\infty$.
The third term in \eqref{Ethnzi}, however, converges to the 
quantity $\frac{1}{S}
\tr\{\vec{\Sigma}_i^+\vec{\Delta}_i\vec{\Delta}_i\herm\vec{\Sigma}_i\pherm)\}$
as $\rho\rightarrow\infty$, where $(\cdot)^+$ denotes pseudo-inverse. 
Now, since $\vec{F}\of{k}\plt$ and $\vec{F}\pli$ are distinct full rank matrices with  
$\tr\{\vec{F}\ofH{k}\plt \vec{F}\of{k}\plt\} = \tr\{\vec{F}\pli\herm \vec{F}\pli\}$, it follows that 
$\vec{\Sigma}_i^+\vec{\Delta}_i \neq \vec{0}$ and hence
$\tr\{\vec{\Sigma}_i^+\vec{\Delta}_i\vec{\Delta}_i\herm\vec{\Sigma}_i\pherm)\} > 0$. 
So there does not exist $C$ such that
$\E\{\norm{\tvec{h}\of{k}\nzpi}^2\} \leq C \rho^{-1}$ for all $\rho>0$.
\end{IEEEproof}

\begin{corollary}
\lemref{lmmse}, and several other results in the paper, 
are stated under prime $N$, arbitrary $\mc{N}\pl$, and $L<N$.
The requirement that $N$ is prime can be relaxed in exchange for 
the following restrictions on $\mc{N}\pl$ and $L$.
\begin{enumerate}
\item 
The set $\mc{N}\pl$ does not form a group with respect to modulo-$N$ 
addition, nor a coset of a subgroup of $\{0,1,\dots,N-1\}$ 
under modulo-$N$ addition. 
\item
The channel length $L$ obeys $L<N/2$.  
\end{enumerate}
\end{corollary}
\begin{IEEEproof}
Throughout the paper, the prime-$N$ property is used only to guarantee 
that certain square submatrices of the $N$-DFT matrix $\vec{F}$ remain
full rank.
When forming these submatrices, we use $S$ row indices from 
$\mc{N}\pl$ (where $\mc{N}\pl\subset \{0,\dots,N-1\}$ and 
$|\mc{N}\pl|=P\geq S$) and $S$ column indices from $\mc{L}_i$ 
(where $\mc{L}_i\subset \{0,\dots,L-1\}$ and $|\mc{L}_i|=S$).
In the case that $N$ is prime, the Chebotarev theorem 
\cite{Stevenhagen:MI:96,Tao:MRL:05}
guarantees that our square submatrix will be full rank,
as discussed in the proof of \lemref{lmmse}. 
However, even when $N$ is not prime, our square submatrix will be 
full rank whenever both $\mc{N}\pl$ and $\mc{L}_i$ do not form groups
with respect to modulo-$N$ addition, nor cosets of subgroups of
$\{0,1,\dots,N-1\}$ w.r.t modulo-$N$ addition \cite[p.491]{Candes:TIT:06}.
These conditions on $\mc{N}\pl$ and $\mc{L}_i$ are ensured by the two 
conditions stated in the corollary.
\end{IEEEproof}

For a given communication scheme, we say that a rate $\mc{R}$ (in bits per 
channel use) is \emph{achievable} if the probability of decoding error can 
be made arbitrarily small at that rate. 
Now, using the bound on the estimation error variance from \lemref{lmmse}, 
we establish that when the true channel support is apriori known at receiver 
(i.e., the support-genie case), the achievable rates satisfy 
$\limit{\rho}{\infty} \frac{\mc{R}(\rho)} {\log \rho} = 1-\frac{P}{N}$, 
where $P \geq S$ denotes the number of pilot tones. 

\begin{lemma} \label{lem:spepro}
Say that $N$ is prime, and that the true channel support is known apriori 
at the receiver for each fading block. 
Then, for any pilot pattern $\mc{N}\pl$ such that $P\geq S$,
the achievable rate of the support-hypothesized estimator-decoder satisfies 
$\limit{\rho}{\infty} \frac{\mc{R}(\rho)}
{\log \rho} = 1-\frac{P}{N}$.
\end{lemma}
\begin{IEEEproof}
The achievable rate of WMD decoding under imperfect channel state 
information (CSI) and Gaussian coding was studied in \cite{Weingarten:TIT:04},
where the rate expressions were obtained under certain restrictions on 
the statistical properties of the imperfect CSI. 
In the support-genie case, our support-hypothesized channel estimator 
satisfies all of the standard requirements in \cite{Weingarten:TIT:04} 
except for time-invariance, since the support varies over the fading blocks.
However, our model does satisfy the alternative ergodic condition in
\cite{Weingarten:TIT:04}. 
To see this, we need to verify that, for any function $f(\cdot)$, we have
$\lim_{ K \rightarrow \infty} \frac{1}{K} \sum_{k=1}^{K} 
f(\vec{y}\da\of{k},\hvec{h}\of{k}\fpikt) 
= \E \big\{ f(\vec{y}\da\of{k},\hvec{h}\of{k}\fpikt) \big\}$, 
using $i_{k,\text{\sf true}}$ to denote the index of the true support 
during the $k^{th}$ fading block, and
$\hvec{h}\of{k}\fpikt \defn \sqrt{N}\vec{F}\ikt\hvec{h}\nzt\of{k}$. 
Let us define $\mc{K}_i = \{k: \mc{L}\of{k}\true = \mc{L}_i\}$  
for $i=1,\dots,M$.
Then it follows that,
\begin{eqnarray}
\lim_{ K \rightarrow \infty} \frac{1}{K} \sum_{k=1}^{K} f(\vec{y}\da\of{k},\hvec{h}\of{k}\fpikt) 
&=& \lim_{K \rightarrow \infty} 
	\frac{1}{K} \sum_{i=1}^M \sum_{k \in \mc{K}_i}  
	f(\vec{y}\da\of{k},\hvec{h}\of{k}\fpi), \\
&=& \sum_{i=1}^M \lim_{K \rightarrow \infty} 
	\frac{|\mc{K}_i|}{K} \frac{1}{|\mc{K}_i|} 
	\sum_{k \in \mc{K}_i} f(\vec{y}\da\of{k},\hvec{h}\of{k}\fpi), \\
&=& \sum_{i=1}^{M} \lambda_i 
	\E \big\{ f(\vec{y}\da\of{k},\hvec{h}\of{k}\fpi) 
	\biggiv \mc{L}\of{k}\true = \mc{L}_i \big\}, 	\label{eq:ergod1}\\
&=& \E\big\{ f(\vec{y}\da\of{k},\hvec{h}\of{k}\fpikt)\big\}.
							\label{eq:ergod2}
\end{eqnarray} 
Hence \cite[Theorem 2]{Weingarten:TIT:04} can be applied to find the 
achievable rates for our decoupled decoding scheme under the support genie.
In particular, by rewriting the data observations from \eqref{edi} as
\begin{eqnarray}
 \vec{y}\of{k}\da
 &=& \sqrt{\rho}\Diag(\vec{x}\of{k}\da)\vec{J}\da \hvec{h}\of{k}\fpikt +
	\vec{e}\of{k}\dikt,
							\label{eq:edi2}
\end{eqnarray}
for effective noise $\vec{e}\of{k}\dikt 
\defn \sqrt{\rho}\Diag(\vec{x}\of{k}\da)\vec{J}\da \tvec{h}\of{k}\fpikt 
+ \vec{v}\of{k}\da$, 
it follows \cite{Weingarten:TIT:04} that the achievable rate 
(in bits per channel use) is
\begin{eqnarray}
\mc{R}(\rho) 
&=& \frac{1}{N}
	\E \Big\{ \log \det \Big[ \vec{I} + \rho\, \vec{C}\dikt^{-1}(\rho)
 	\Diag\big(\vec{J}\da\hvec{h}\of{k}\fpikt\big) \vec{R}\da  
 	\Diag\big(\vec{J}\da\hvec{h}\of{k}\fpikt\big)\herm \Big] \Big\},   
							\label{eq:ar} 
\end{eqnarray}
where $\vec{C}\di(\rho) \defn \cov\{\vec{e}\of{k}\di\}$ for 
$\vec{e}\of{k}\di$ defined in \eqref{edi}.
Similar to \eqref{ergod1}-\eqref{ergod2}, we can rewrite \eqref{ar} as
\begin{eqnarray}
\mc{R}(\rho) 
&=& \frac{1}{N}\sum_{i=1}^{M} \lambda_i 
	\underbrace{ \E \bigg\{  
	\log \det \Big[ \vec{I} + \rho \vec{C}\di^{-1}(\rho) 
 	\Diag\big(\vec{J}\da\hvec{h}\of{k}\fpi\big) \vec{R}\da  
	\Diag\big(\vec{J}\da\hvec{h}\of{k}\fpi\big)\herm \Big] 
	\Biggiv \mc{L}\of{k}\true = \mc{L}_i \bigg\} 
	}_{\displaystyle \defn\mc{R}_i(\rho)} . \quad	\label{eq:ar1} 
\end{eqnarray}
When $\mc{L}\of{k}\true = \mc{L}_i$, \lemref{lmmse} specifies that
there exists some constant $C$ such that
$\E\{\norm{\tvec{h}\of{k}\nzpi}^2\} \leq C\rho^{-1}$ for all $\rho$.
In this case, the eigenvalues of $\vec{C}\di(\rho)$ will be positive
and bounded from above for all $\rho$, and thus eigenvalues of 
$\vec{C}^{-1}\di(\rho)$ will be positive and bounded from below 
for all $\rho$.
Thus, using a standard high-SNR analysis 
(see, e.g., \cite{Kannu:TIT:10} for details), 
$\lim_{\rho\rightarrow\infty}\frac{\mc{R}_i(\rho)}{\log \rho} = 1-\frac{P}{N}$
for any $i$, from which the stated result of this lemma follows.
\end{IEEEproof}

In \cite{Vikalo:TSP:04}, it has been shown that, for $L$-length
non-sparse channels, PAT can be designed to achieve data rates that satisfy 
$\lim_{\rho\rightarrow\infty}\frac{\mc{R}(\rho)}{\log \rho} = 1-\frac{P}{N}$, for $P \geq L$. 
Our \lemref{spepro} can be interpreted as an extension of the result 
from \cite{Vikalo:TSP:04} to $L$-length $S$-sparse channels with known
support.

\section{Channel-Support Decoding}

In summary, the PAT scheme of \secref{patdefn} and the decoupled decoder of
\secref{decoupled} will suffice for spectral efficient communication over
the sparse frequency-selective block-fading channel \emph{if} we can 
establish a reliable means of determining the correct support (i.e., 
$\vec{i}$ such that $\vec{\mc{L}}_{\vec{i}}=\vec{\mc{L}}\true$).
In this section, we consider schemes for reliably decoding the channel support
of each block.

\subsection{Data-Aided Support Decoding}		\label{sec:dasd}
In this section, we show that, with prime $N$, the
pilot aided transmission (PAT) scheme defined in \secref{patdefn} 
is spectrally efficient for the sparse frequency-selective block-fading 
channel.
In other words, when the $L$-length channel is $S$-sparse, it is sufficient 
to sacrifice only $S$ signal-space dimensions to maintain an achievable rate
that grows at the same rate as channel capacity in the high-SNR regime.
To show this, we construct a so-called \emph{data-aided support decoder} 
(DASD) that leverages certain error-detecting capabilities in the 
codebook $\code$. 
We first describe the error detection mechanism and later 
propose a procedure for channel support decoding. 

\textb{
In our DASD scheme, we attach error detection parity bits, which we refer to
as cyclic redundancy check (CRC) bits, to the information bits \emph{prior} to 
the channel-coding operation. Attaching parity bits to the information bits is a commonly used 
mechanism to identify the decoding errors at the receiver \cite{Wicker:Book:95}.  
Let us denote the information bit rate as $R$, and the CRC bit rate as 
$\delta$, both in units of bits-per-channel-use. 
Then, over $m=KN$ channel uses, we use a total\footnote{
  For ease of presentation, we have ignored the flooring 
  $\lfloor mR \rfloor$ and $\lfloor m\delta \rfloor$ and the flooring 
  error can be made negligible by choosing a large $m$.} 
of $mR$ bits for information and a total of $m\delta$ bits for CRC. 
Let $\mu(\cdot)$ denote the function which specifies the $m\delta$ parity bits for every set of $mR$ information bits. 
Specifically, $\mu: \{ 1,\dots,2^{mR} \} \rightarrow \{1,\dots,2^{m\delta}\}$ 
is a ``binning function'' mapping information bits to corresponding CRC bits,
so that,
for the information message $\vec{w}$, the corresponding CRC bits
are $\vec{u}=\mu(\vec{w})$. 
Such $\vec{u}$ is sometimes referred to as the ``auxiliary check message.'' 
The channel-encoder then maps the ``composite message'' $(\vec{w},\vec{u})$, 
containing $m(R+\delta)$ bits, to one of the $2^{m(R+\delta)}$ codewords 
in the codebook $\code$. 
(See \secref{patdefn} for details on the codebook.) 
For clarity, we use ``message'' when referring to channel-coder inputs, 
and ``codeword'' when referring to channel-coder outputs.}

The DASD support decoding procedure is defined as follows.
\begin{quote}
For each hypothesis of support index
$\vec{i}=(i_1,\dots,i_K) \in \{1,\dots,M\}^K$, 
\begin{enumerate}
\item Compute conditional channel estimates $\{\hvec{h}\of{k}\fpik\}_{k=1}^K$ 
	and $\{\vec{\Sigma}\fpik\}_{k=1}^K$ 
      	using \eqref{hnzpi}-\eqref{Signzpi} 
	with $\hvec{h}\of{k}\fpik=\sqrt{N}\vec{F}\ik\hvec{h}\of{k}\nzpik$ 
	and $\vec{\Sigma}\fpik=N\vec{F}\ik\vec{\Sigma}\nzpik\vec{F}\ik\herm$. 
\item Compute the WMD codeword estimate $\hvec{x}\dii$ 
	according to \eqref{deco}.
\item From the codeword $\hvec{x}\dii$, recover the corresponding composite message $(\hvec{w}\ii,\hvec{u}\ii)$. 
\item Perform error detection on $(\hvec{w}\ii,\hvec{u}\ii)$, i.e., check if $\mu(\hvec{w}\ii) \neq \hvec{u}\ii$. 
\item If no error is detected or there are no more hypotheses to consider, 
	stop and declare the decoded message as $\hvec{w}\ii$, 
	else continue with the next hypothesis $\vec{i}$. 
\end{enumerate} 
\end{quote}
The asymptotic performance of DASD is characterized by the following theorem.

\begin{theorem}
For the $S$-sparse frequency-selective $N$-block-fading channel with 
prime $N$, 
the previously defined PAT scheme, when used with $S$ pilots and DASD, 
yields an achievable rate $\mc{R}\dasd(\rho)$ that obeys 
$\lim_{\rho\rightarrow\infty} \frac{\mc{R}\dasd(\rho)}{\log\rho} 
= 1-\frac{S}{N}$.
Hence, PAT is spectrally efficient for this channel.
\end{theorem}

\begin{IEEEproof}
In our proof, instead of considering a specific binning function $\mu(\cdot)$, we consider the error performance 
averaged over all possible random binning assignments and establish that the average error approaches zero. 
For a given support hypothesis $\vec{\mc{L}}\ii$, the DASD
computes the support-conditional channel estimate and the corresponding
WMD codeword estimate from which the composite message bits are obtained, 
which we write as $(\hvec{w}\ii,\hvec{u}\ii)$.
There are two situations under which the DASD terminates,
producing the final estimate $\hvec{w}\dasd=\hvec{w}\ii$:
i) when $\vec{i}\neq\vec{i}\last$ and $\mu(\hvec{w}\ii) = \hvec{u}\ii$, or
ii) when $\vec{i}=\vec{i}\last$.
Here we use $\vec{i}\last$ to denote the last of the $M^K$ hypotheses.
Note that, in all other cases, 
an error is detected, and the DASD 
continues under a different hypothesis $\vec{\mc{L}}\iip$.

We now upper bound the probability that the DASD
infers the wrong information bits, i.e., that $\hvec{w}\dasd\neq \vec{w}$.
Say that $\vec{i}\stp$ denotes the value of $\vec{i}$ used to 
produce $\hvec{w}\dasd$, i.e., $\hvec{w}\dasd = \hvec{w}\iis$.
Notice that either 
1) $\vec{i}\stp=\vec{i}\true$ or
2) $\vec{i}\stp\neq\vec{i}\true$. 
In the latter case, the support detector fails to detect the true support 
when either 
2a) $\vec{i}\stp\neq\vec{i}\last$ and $\mu(\hvec{w}\iis) = \hvec{u}\iis$,
where the error was missed, or
2b) $\vec{i}\stp=\vec{i}\last$.
Finally, notice that, if event 2b occurs, the DASD must have (falsely)
detected an error under the true support hypothesis, i.e., 
$\mu(\hvec{w}\iit) \neq \hvec{u}\iit$.
Thus we can partition the error event $\hvec{w}\iis\neq \vec{w}$ into three 
mutually exclusive events:
\begin{enumerate}
\item[E1)] $\vec{i}\stp=\vec{i}\true$ and $\hvec{w}\iis\neq \vec{w}$,
\item[E2)] $\vec{i}\stp=\vec{i}\last\neq\vec{i}\true$ and both 
	$\mu(\hvec{w}\iit) \neq \hvec{u}\iit$ and $\hvec{w}\iis\neq \vec{w}$.
\item[E3)] $\exists \vec{i}\stp\notin\{\vec{i}\true,\vec{i}\last\} 
	\st \text{both~} \mu(\hvec{w}\iis) = \hvec{u}\iis 
	\text{~and~} \hvec{w}\iis\neq \vec{w}$.
\end{enumerate}
We now analyze each of these three events.

Notice that E1 is the event of a data-decoding error under the correct 
support hypothesis (i.e., $\hvec{w}\iit\neq \vec{w}$).
We recall that the correct-support-hypothesis case was analyzed in 
\secref{decoupled}, under which PAT with decoupled decoding was found 
to be spectrally efficient, having an achievable rate $\mc{R}$ that obeys
$\lim_{\rho\rightarrow\infty}\frac{\mc{R}(\rho)}{\log \rho} = 1-\frac{S}{N}$.
Thus, the probability of E1 can be made arbitrarily small
for any rates $R$ and $\delta$ such that $R+\delta\leq \mc{R}$. 

E2 characterizes the event in which 
the true support is falsely discarded and data-decoding error 
results later (under an incorrect support hypothesis).
Recall that, when the support hypothesis is incorrect, we cannot guarantee
a low probability of 
data-decoding error when communicating at rates that scale as 
$(1-\frac{S}{N})\log\rho$.
The key, then, is to make the support-error probability small.
Towards this aim, we bound E2 as follows:
\begin{eqnarray}
  \Pr\{\text{E2}\}
  &=& \Pr\{ \mu(\hvec{w}\iit) \neq \hvec{u}\iit 
  	\text{~and~} \hvec{w}\iis\neq \vec{w}\} \\
  &\leq& \Pr\{ \mu(\hvec{w}\iit) \neq \hvec{u}\iit \} \\
  &=& \Pr\{ \mu(\hvec{w}\iit) \neq \hvec{u}\iit \giv \hvec{w}\iit = \vec{w}\} 
  	\Pr\{ \hvec{w}\iit = \vec{w} \}
	\nonumber\\&&\mbox{}
  	+ \Pr\{ \mu(\hvec{w}\iit) \neq \hvec{u}\iit 
		\giv \hvec{w}\iit\neq \vec{w}\} 
  	\Pr\{ \hvec{w}\iit \neq \vec{w} \} \\
  &\leq& \Pr\{ \mu(\hvec{w}\iit) \neq \hvec{u}\iit \giv \hvec{w}\iit = \vec{w}\}
  	+ \Pr\{ \hvec{w}\iit \neq \vec{w} \} \\
  &=& \Pr\{ \vec{u} \neq \hvec{u}\iit \} + \Pr\{ \hvec{w}\iit \neq \vec{w} \}.
  							\label{eq:E2}
\end{eqnarray}
Thus, the probability of E2 can be upper bounded by the probability of
decoding error under the correct support-hypothesis, which (like 
$\Pr\{\text{E1}\}$) can be made arbitrarily small for any achievable rate. 

E3 describes the event that both the detection of a support-error is missed 
and a data-decoding error results.
Like with E2, the probability of data-decoding cannot be made arbitrarily
small under an incorrect support hypothesis, and so we hope that the
false alarm error is small.
Towards this aim, we begin by upper bounding the probability of the event E3 
as follows:
\begin{eqnarray}
  \lefteqn{ \Pr\{\text{E3}\} }\nonumber\\
  &=& \Pr\big\{ \exists~\vec{i}\stp\notin\{\vec{i}\true,\vec{i}\last\} 
        \st \mu(\hvec{w}\iis) = \hvec{u}\iis 
	\biggiv \hvec{w}\iis\neq \vec{w}\big\} \Pr\{\hvec{w}\iis\neq \vec{w}\} 
		\quad \\
  &\leq& \Pr\big\{ \exists~\vec{i}\notin\{\vec{i}\true,\vec{i}\last\} 
        \st \mu(\hvec{w}\ii) = \hvec{u}\ii 
	\biggiv \hvec{w}\ii\neq \vec{w}\big\} \\
  &\leq& \Pr\{ \exists~\vec{i}\neq \vec{i}\true 
        \st \mu(\hvec{w}\ii) = \hvec{u}\ii \giv \hvec{w}\ii\neq \vec{w} \} \\
  &\leq& \sum_{\vec{i}\neq\vec{i}\true} 
  	\Pr\{\mu(\hvec{w}\ii) = \hvec{u}\ii \giv \hvec{w}\ii\neq \vec{w} \} 
							\label{eq:ub}
\end{eqnarray}
where we used the union bound in \eqref{ub}.
Now, to find the probability of missing a support-error, we assume that,
when $\hvec{w}\ii\neq \vec{w}$, the auxiliary check estimate $\mu(\hvec{w}\ii)$ 
is uniformly distributed over all possibilities of $\vec{u}$. 
This can be justified by letting the function $\mu$ be 
constructed by a random binning assignment of the codewords onto 
$2^{m\delta}$ bins, and averaging over the ensemble of random binning 
assignments \cite{Elgamal:TIT:06}.  
In this case, for any $\vec{i} \neq \vec{i}\true$, the probability of 
missing the detection of a support-error becomes
\begin{eqnarray}
 \Pr\{\mu(\hvec{w}\ii) = \hvec{u}\ii 
          \giv \vec{i}\neq\vec{i}\true, \hvec{w}\ii\neq \vec{w}\}
 &=& \frac{1}{2^{m\delta}},
\end{eqnarray} 
so that
\begin{eqnarray}
 \Pr\{\text{E3}\} 
 &\leq&  \frac{M^K}{2^{m\delta}} 
 ~=~  \frac{M^K}{2^{KN\delta}} 
 ~=~  \bigg( \frac{M}{2^{N\delta}} \bigg)^K .
\end{eqnarray}
So, when $\delta > \frac{ \log M } {N}$, by choosing $K$ 
large enough, we can make $\Pr\{\text{E3}\}$ averaged over all the 
random binning CRC assignments arbitrarily small. 
This implies that there exists a binning function $\tilde{\mu}$ for which 
$\Pr\{\text{E3}\}$ can be made arbitrarily small. 

Notice that the rate $\delta$ sacrificed to make $\Pr\{\text{E3}\}$ 
arbitrarily small does not grow with SNR $\rho$. 
As long as we choose the SNR-dependent information rate 
$R(\rho) \leq \mc{R}(\rho) - \delta$, where $\mc{R}(\rho)$ is an 
achievable rate for the sparse channel with known support described 
in \lemref{spepro}, we can construct a codebook that guarantees arbitrarily 
small values for $\Pr\{\text{E1}\}+\Pr\{\text{E2}\}$. 
This codebook, when used in conjunction with the binning function 
$\tilde{\mu}$, ensures that $\Pr\{\hvec{w}\dasd\neq \vec{w}\}
=\Pr\{\text{E1}\}+\Pr\{\text{E2}\}+\Pr\{\text{E3}\}$ 
can be made arbitrarily small. 
Since $\delta$ is fixed with respect to SNR $\rho$, the information rate 
of DASD satisfies 
$\lim_{\rho\rightarrow\infty} \frac{R(\rho)}{\log \rho} = 1-\frac{S}{N}$.
\end{IEEEproof}

As we have seen, the DASD achieves the optimal pre-log factor,
albeit at complexity\footnote{
  Note that the term to the right of the sum in the WMD decoder metric 
  \eqref{deco} must be computed for every triple $(i,k,\vec{x}\of{k}\da)$, 
  where the complexity of each computation is $\mc{O}(N^2)$.  Subsequently,
  these terms must be summed for each of $M^K$ support-vector hypotheses.}
$\mc{O}(|\code| M^K + |\code| M K N^2)$,
which may be larger than that of the optimal decoder specified in \lemref{optd}.
In fact, we do not propose DASD for practical use, but rather as a 
constructive means of proving the achievability of the optimal pre-log factor,
since the optimal decoder is difficult to analyze directly.
In the next section, we present a simpler suboptimal decoding scheme
that also has performance guarantees.

\subsection{Pilot-Aided Support Decoding}		\label{sec:pasd}
In this section, we propose a \emph{pilot-aided support decoder} (PASD) 
with complexity\footnote{
  As described below, 
  for support estimation, $K$ instances of $\hat{\imath}\pl\of{k}$ must 
  be computed, each with complexity $\mc{O}(M P^2)$.
  Then, for (support-conditional) WMD decoding, $|\code| K$ instances of the 
  term after the sum in \eqref{deco} must be computed, each with 
  complexity $\mc{O}(N^2)$.}
$\mc{O}(|\code| K N^2 + K M P^2)$, which is significantly less complex than 
both DASD and the optimal decoder in \lemref{optd}.
Since only pilots are used to infer the channel support, 
the complexity of support estimation grows linearly in $K$.
PASD, however, requires one additional pilot dimension relative to DASD
(i.e., $P=S+1$) and is only \emph{asymptotically reliable} 
(i.e., the probability of support-detection error vanishes as 
$\rho\rightarrow\infty$ but is not guaranteed to be arbitrarily small 
at any finite $\rho$) unless the channel support $\mc{L}\of{k}$ is
fixed over fading blocks $k\in\{1,\dots,K\}$.

\subsubsection{Pilot-Aided Support Estimation}		\label{sec:pase}
We now present an asymptotically reliable method to infer the channel 
support $\vec{\mc{L}}$ that requires only $P=S+1$ pilots per fading block. 
For this, we use the following normalized pilot observations:
\begin{eqnarray}
  \vec{z}\of{k}\pl
  &\defn& \textstyle
  	\frac{1}{\sqrt{\rho N}} \Diag(\vec{x}\pl^*) \vec{y}\of{k}\pl 
  ~=~ \vec{F}\of{k}\plt \vec{h}\of{k}\nz 
  	+ \frac{1}{\sqrt{\rho N}} \vec{\nu}\of{k}\pl ,
\end{eqnarray}
where $\vec{\nu}\of{k}\pl \sim \mc{CN}(\vec{0},\vec{I})$ due to the
constant-modulus assumption on the pilots.
Recalling that $\vec{F}\of{k}\plt$ is constructed from 
rows $\mc{N}\pl$ and columns $\mc{L}\of{k}\true$ of $\vec{F}$,
and that $\vec{F}\pli$ is constructed from 
rows $\mc{N}\pl$ and columns $\mc{L}_i$ of $\vec{F}$, 
we henceforth use $\vec{\Pi}\pli \defn 
\vec{F}\pli(\vec{F}\pli\herm\vec{F}\pli)^{-1}\vec{F}\pli\herm$
to denote the matrix that projects onto the column space of $\vec{F}\pli$,
and $\vec{\Pi}\pli^\perp \defn \vec{I}-\vec{\Pi}\pli$ to denote its 
orthogonal complement.

The pilot-aided support estimator (PASE) infers the support index
as that which minimizes the energy of the projection error 
$\vec{e}\of{k}\pli$:
\begin{eqnarray}
 \hat{\imath}\of{k}\pl
 &\defn& \argmin_{i\in\{1,\dots,M\}} \norm{\vec{e}\of{k}\pli}^2
 \text{~~for~~}
 \vec{e}\of{k}\pli 
 ~\defn~\vec{\Pi}\pli^\perp \vec{z}\of{k}\pl 		\label{eq:PASE}
\end{eqnarray}
Clearly, the complexity of PASE is proportional to 
$M={L \choose S}=\mc{O}\big((L/S)^S\big)$.
Thus, while the complexity of PASE is much less than the DASD 
proposed in \secref{dasd}, we note that its complexity may be significantly
larger than classical compressive sensing algorithms like basis pursuit,
whose complexity is polynomial in $L$ \cite{Donoho:TIT:06b}.

\begin{theorem}         \label{thm:ase}
For the $S$-sparse frequency-selective $N$-block-fading channel with 
prime $N$, and the previously defined PAT scheme with $P\geq S+1$
arbitrarily placed pilots, the probability of PASE support-detection error
vanishes as $\rho\rightarrow\infty$.
\end{theorem}

\begin{IEEEproof}
We first note that, due to the Chebotarev theorem 
\cite{Stevenhagen:MI:96,Tao:MRL:05},
each $\vec{F}\pli\in\Complex^{P\times S}$ is full rank 
when $N$ is prime and $P\geq S+1$.
Also, each column $\vec{f}$ of $\vec{F}\pli$ is linearly independent of all 
columns in $\vec{F}\plj\big|_{j\neq i}$ that are not equal to $\vec{f}$.
Thus, each $\vec{F}\pli$ defines a \textit{unique} column space. 
We note that this property does not hold when $P=S$. 

A PASE support-detection error results when 
$\exists i\neq i\of{k}\true \st 
\norm{\vec{e}\of{k}\pli}^2 < \norm{\vec{e}\of{k}\plt}^2$.
The probability of this event can be upper bounded as follows,
\begin{eqnarray}
 \lefteqn{
 \Pr\big\{ \exists i\neq i\of{k}\true \st 
 	\norm{\vec{e}\of{k}\pli}^2 < \norm{\vec{e}\of{k}\plt}^2 \big\}
 }\nonumber\\
 &\leq& \sum_{i\neq i\of{k}\true} \Pr\big\{ 
 	\norm{\vec{e}\of{k}\pli}^2 < \norm{\vec{e}\of{k}\plt}^2 \big\} 
							\label{eq:ub2}\\
 &=& \sum_{i\neq i\of{k}\true} \Pr\big\{ \textstyle
 	\norm{ \vec{\Pi}\pli^\perp\vec{F}\of{k}\plt \vec{h}\of{k}\nz
		+ \frac{1}{\sqrt{\rho N}} 
		\vec{\Pi}\pli^\perp \vec{\nu}\of{k}\pl }
 	< \frac{1}{\sqrt{\rho N}} \norm{
		\vec{\Pi}\ofP{k}\plt \vec{\nu}\of{k}\pl} \big\} \\
&\leq&  \sum_{i\neq i\of{k}\true} \Pr\big\{ \textstyle
 	\norm{ \vec{\Pi}\pli^\perp\vec{F}\of{k}\plt \vec{h}\of{k}\nz}
		- \norm{ \frac{1}{\sqrt{\rho N}} 
		\vec{\Pi}\pli^\perp \vec{\nu}\of{k}\pl }
 	< \frac{1}{\sqrt{\rho N}} \norm{
		\vec{\Pi}\ofP{k}\plt \vec{\nu}\of{k}\pl} \big\} \label{eq:ub3} \\
&=&  \sum_{i\neq i\of{k}\true} \Pr\big\{ \textstyle
 	\norm{ \vec{\Pi}\pli^\perp\vec{F}\of{k}\plt \vec{h}\of{k}\nz}
		 	<  \frac{1}{\sqrt{\rho N}}  \norm{ 
		\vec{\Pi}\pli^\perp \vec{\nu}\of{k}\pl }
		  + \frac{1}{\sqrt{\rho N}} \norm{
		\vec{\Pi}\ofP{k}\plt \vec{\nu}\of{k}\pl} \big\} \label{eq:ub4}\\
 &\leq& 
\sum_{i\neq i\of{k}\true} \Pr\big\{ \textstyle
 	\norm{ \vec{\Pi}\pli^\perp\vec{F}\of{k}\plt \vec{h}\of{k}\nz}
      < \frac{2}{\sqrt{\rho N}} \norm{\vec{\nu}\of{k}\pl} \big\},\label{eq:pase}
\end{eqnarray}
where the probability of error in \eqref{ub3} was upper-bounded by 
making the left side of the inequality smaller via 
$\|\vec{x}\| - \|\vec{y}\| \leq \|\vec{x} + \vec{y}\|$.
The upper bound \eqref{pase} follows from 
$\norm{ \vec{\Pi}\pli^\perp \vec{\nu}\of{k}\pl } \leq \norm{
\vec{\nu}\of{k}\pl } $ and 
$  \norm{\vec{\Pi}\ofP{k}\plt \vec{\nu}\of{k}\pl} \leq \norm{
\vec{\nu}\of{k}\pl }$, which hold because
$\vec{\Pi}\pli^\perp$ and $\vec{\Pi}\plt\ofP{k}$ are projection matrices. 
Taking the SVD $\vec{\Pi}\pli^\perp\vec{F}\of{k}\plt
= \vec{U}_i\of{k}\vec{\Sigma}_i\of{k}\vec{V}_i\ofH{k}$ 
and defining $\vec{g}_i\of{k}\defn\sqrt{S}\vec{V}_i\ofH{k}\vec{h}\of{k}\nz
\sim \mc{CN}(\vec{0},\vec{I})$, we can rewrite \eqref{pase} as
follows and upper bound further:
\begin{eqnarray}
 \lefteqn{
 \Pr\big\{ \exists i\neq i\of{k}\true \st 
 	\norm{\vec{e}\of{k}\pli}^2 < \norm{\vec{e}\of{k}\plt}^2 \big\}
 }\nonumber\\
 &\leq& \sum_{i\neq i\of{k}\true} \Pr\big\{ \textstyle
 	\norm{ \vec{\Sigma}_i\of{k}\vec{g}_i\of{k} }^2
	< \frac{4S}{\rho N} \norm{\vec{\nu}\of{k}\pl}^2 \big\} \\
 &\leq& \sum_{i\neq i\of{k}\true} \Pr\big\{ \textstyle
 	(\sigma_{i,0}\of{k})^2 |g_{i,0}\of{k}|^2
	< \frac{4S}{\rho N} \norm{\vec{\nu}\of{k}\pl}^2 \big\} \\
 &\leq& \sum_{i\neq i\of{k}\true} \Pr\big\{ \textstyle
 	(\sigma_{i,0}\of{\min})^2 |g_{i,0}\of{k}|^2
	< \frac{4S}{\rho N} \norm{\vec{\nu}\of{k}\pl}^2 \big\} \\
 &=& \sum_{i\neq i\of{k}\true} \Pr\left\{
 	\frac{|g_{i,0}\of{k}|^2}{ \norm{\vec{\nu}\of{k}\pl}^2}
	< \frac{4S}{(\sigma_{i,0}\of{\min})^2 \rho N} \right\} .	\label{eq:F}
\end{eqnarray}
Above, $\sigma_{i,0}\of{k}$ denotes the largest singular value in 
$\vec{\Sigma}_i\of{k}$ and  $\sigma_{i,0}\of{\min} \defn \min_{k}\sigma_{i,0}\of{k}$.
Notice that at least one of the columns of  $\vec{F}\of{k}\plt$  lies outside 
the column space of  $\vec{F}\pli$. The projection of those columns onto the subspace orthogonal to
the column space of $\vec{F}\pli$ will be non-zero implying that $\vec{\Pi}\pli^\perp\vec{F}\of{k}\plt$ is not 
identical to $\vec{0}$ and hence the largest singular value $\sigma_{i,0}\of{k} > 0, \forall k$. 
Since $g_{i,0}\of{k}\sim\mc{CN}(0,1)$ is independent of 
$\vec{\nu}\of{k}\pl\sim\mc{CN}(0,\vec{I})$, 
the random variable $F_i\of{k} \defn 
|g_{i,0}\of{k}|^2/\norm{\vec{\nu}\of{k}\pl}^2$
is F-distributed with parameters $(2,2P)$.
Since the cumulative distribution function (cdf) of an F-distributed
random variable vanishes as its argument 
(in this case, $\frac{4S}{(\sigma_{i,0}\of{\min})^2 \rho N}$) approaches zero,
the probability of a PASE error vanishes as $\rho\rightarrow\infty$.
\end{IEEEproof}

We now make a few comments about \thmref{ase}.
To perfectly recover any \emph{arbitrary deterministic} 
$S$-sparse impulse response from noise-free frequency-domain samples,
\cite{Candes:TIT:06} established that $2S$ pilot tones are 
both necessary and sufficient. 
In contrast, to perfectly recover an $S$-sparse \emph{probabilistic}
Rayleigh-fading impulse response, \thmref{ase} establishes that $S+1$ 
noise-free pilot observations suffice \emph{with probability one}. 
In particular, the condition $P \geq S+1$ ensures that the set of 
$\vec{h}\of{k}$ that cannot be recovered by the PASE support detector has 
probability $0$ with respect to the Gaussian distribution on the nonzero 
entries of $\vec{h}\of{k}$. 
To see this, notice that 
$\text{rank}(\vec{F}\pli) = \text{rank}(\vec{F}\plj) = S$, but also that 
$\text{range}(\vec{F}\pli) = \text{range}(\vec{F}\plj)$ only if $i = j$. 
In particular, if $i \neq j$, then 
$\text{dim}\{\text{range}(\vec{F}\pli) \cap \text{range}(\vec{F}\plj)\} 
= S-1$.
This implies that the set of vectors $\vec{h}\nz \in \Complex^S$ for which 
$\vec{F}\pli \vec{h}\nz$ is in the range space of $\vec{F}\plj$ has measure 
zero with respect to any continuous distribution on $\vec{h}\nz$. 
Similar results on the recovery of probabilistic sparse signals have also 
appeared in \cite{Tropp:ACHA:08}.
 
\subsubsection{Pilot-Aided Support Decoding}
For pilot-aided support decoding, we assume that the 
transmitter uses the PAT scheme defined in \secref{patdefn}
with $P=S+1$ pilots and prime $N$.
At the receiver, the PASE scheme described in the previous section is 
used to estimate the sparse channel support and, based on this 
estimate, support-conditional channel estimation and decoupled data decoding
are performed as described in \secref{decoupled}.

We now study the $\epsilon$-achievable rate of PAT with PASD.
For some $\epsilon > 0$ and SNR $\rho$,
let $\mc{R}_{\epsilon}(\rho)$ denote the information rate for which the 
probability of decoding error can be made less than $\epsilon$.
\lemref{pasd} characterizes $\mc{R}_{\epsilon}(\rho)$ for PAT with PASD.

\begin{lemma}						\label{lem:pasd}
For the $S$-sparse frequency-selective $N$-block-fading channel with
prime $N$, the previously defined PAT scheme, when used with $S+1$ pilots 
and PASD, yields an $\epsilon$-achievable rate $\mc{R}\pasd_{\epsilon}$ 
that, for any $\epsilon > 0$, obeys $\lim_{\rho \rightarrow \infty} 
\frac{\mc{R}\pasd_{\epsilon}(\rho)}{\log \rho} = 1-\frac{S+1}{N}$.
\end{lemma}
\begin{IEEEproof}
From \thmref{ase} we know that, under the conditions stated in the lemma,
there exists, for any $\epsilon > 0$, an SNR $\rho_\epsilon$ 
above which the error of PASE is less than $\epsilon/2$.
In the case that the support hypothesis is correct, the channel 
estimation and decoupled decoding of \secref{decoupled} allow for the design of 
a codebook $\code_{\rho,\epsilon}$ that 
guarantees data decoding with error probability less than $\epsilon/2$ 
at SNR $\rho$.
Furthermore, from \lemref{spepro}, this codebook can be designed with a
rate $R_{\epsilon}(\rho)$ such that
$\lim_{\rho\rightarrow\infty} \frac{R_{\epsilon}(\rho)}{\log\rho}
=1-\frac{S+1}{N}$. 
Putting these together, we obtain the result of the lemma.
\end{IEEEproof}

We note that, for any given finite SNR $\rho$, it is not possible to make
$\epsilon$, the PASD error probability, arbitrarily small.
Thus, the achievable rate $\mc{R}(\rho)$ of PAT with PASD equals zero for
any finite $\rho$. 
This behavior contrasts that of PAT with DASD, which had positive 
achievable rate for all $\rho > 0$.

Recall that, with the sparse block-fading channel model 
assumed throughout the paper, the channel support 
$\mc{L}\of{k}$ changes independently over fading blocks $k$.
We now consider a variation of this channel for which the 
support does not change\footnote{
  Although the support $\mc{L}\of{k}$ remains fixed over $k$, the 
  nonzero channel taps $\vec{h}\of{k}\nz$ still vary 
  independently over $k$.}
over $k$.
For this fixed-support channel, it is possible to modify PASE so that it 
recovers the support $\vec{\mc{L}}$ with an arbitrarily small probability 
of error at any SNR $\rho > 0$, leading to the 
following corollary of \lemref{pasd}.

\begin{corollary}					\label{cor:pasd}
For the $S$-sparse frequency-selective $N$-block-fading channel with
prime $N$ and a support $\{\mc{L}\of{k}\}_{k=1}^K$ that is 
constant over the fading block index $k$, the previously defined PAT 
scheme, when used with $S+1$ pilots and PASD, yields an achievable 
rate $\mc{R}\pasd$ that obeys $\lim_{\rho \rightarrow \infty} 
\frac{\mc{R}\pasd(\rho)}{\log \rho} = 1-\frac{S+1}{N}$.
\end{corollary}
\begin{IEEEproof}
For this channel, we use PASE with the metric 
$\frac{1}{K}\sum_{k=1}^K \norm{\vec{e}\of{k}\pli}^2$ 
in place of the metric $\norm{\vec{e}\of{k}\pli}^2$ from \eqref{PASE}.
With this modification, we obtain an error probability upper-bound
analogous to \eqref{F}, but where the F-distributed random variable 
has parameters $(2K,2K(S+1))$.
In particular,
\begin{eqnarray}
 \lefteqn{
 \Pr\left\{ \exists i\neq i\true \st 
 	\sum_{k=1}^K\norm{\vec{e}\of{k}\pli}^2 
	< \sum_{k=1}^K\norm{\vec{e}\of{k}\plt}^2 \right\}
 }\nonumber\\
 &\leq& \sum_{i\neq i\true} \Pr\left\{
 	\frac{\sum_{k=1}^K|g_{i,0}\of{k}|^2}
		{\sum_{k=1}^K\norm{\vec{\nu}\of{k}\pl}^2}
	< \frac{S}{(\sigma_{i,0}\of{\min})^2 \rho N} \right\} .	\label{eq:F1}
\end{eqnarray}
For an F-distributed random variable with parameters $(2K,2K(S+1))$, 
the value of the cdf at any fixed point decreases with $K$.
Thus, by choosing a suitably large $K$, we can make the PASE 
support-detection error arbitrarily small at any SNR $\rho>0$.
The result of this lemma then follows from \lemref{spepro}. 
\end{IEEEproof}


\section{Conclusion}

In this paper, we considered the problem of communicating reliably over 
frequency-selective block-fading channels whose impulse responses are 
sparse and whose realizations are unknown to both transmitter and receiver,
but whose statistics are known.
In particular, we considered discrete-time channel impulse responses with 
length $L$ and sparsity exactly $S\leq L$, whose support and coefficients 
remain fixed over blocks of $N>L$ channel uses but change independently from 
block to block. 

Assuming that the non-zero coefficients and noise are both Gaussian, 
we first established that the ergodic noncoherent channel capacity 
$\mc{C}\sparse(\rho)$ obeys 
$\lim_{\rho \rightarrow \infty} \frac{\mc{C}\sparse(\rho)}{\log_2 \rho} 
= 1-\frac{S}{N}$ for any $L$.
Then, we shifted our focus to pilot-aided transmission (PAT), where we 
constructed a PAT scheme and a so-called data-aided support decoder (DASD)
that together enable communication with arbitrarily small error 
probability using only $S$ pilots per fading block.
Furthermore, we showed that the achievable rate $\mc{R}\dasd(\rho)$ of this 
pair exhibits the optimal pre-log factor, i.e.,
$\lim_{\rho \rightarrow \infty} \frac{\mc{R}\dasd(\rho)}{\log_2 \rho} 
= 1-\frac{S}{N}$. 
The use of $S$ pilots can be contrasted with ``compressed OFDM channel 
sensing,'' for which $\mc{O}(S \ln^5 N)$ pilots are known to suffice
for accurate channel estimation (with high probability) in the presence
of noise, and for which $2S$ pilots are known to be necessary and sufficient 
for perfect channel estimation in the absence of noise.

Due to the complexity of DASD,
we also proposed a simpler pilot-aided support decoder (PASD) 
that requires only $S+1$ pilots per fading block. 
For PASD, the $\epsilon$-achievable rate $\mc{R}\pasd_\epsilon(\rho)$ 
obeys, for any $\epsilon>0$, 
$\lim_{\rho \rightarrow \infty} \frac{\mc{R}\pasd_\epsilon(\rho)}{\log_2 \rho} 
= 1-\frac{S+1}{N}$ with the previously considered 
channel, and its achievable rate $\mc{R}\pasd(\rho)$ obeys
$\lim_{\rho \rightarrow \infty} \frac{\mc{R}\pasd(\rho)}{\log_2 \rho} 
= 1-\frac{S+1}{N}$ when the sparsity pattern of the block-fading channel 
remains fixed over fading blocks.
We note that, in recent work \cite{Schniter:ASIL:10,Schniter:PHYCOM:11}, 
the authors have proposed a loopy belief propagation based joint channel 
estimation and decoding scheme, with complexity $\mc{O}(KLN)$,
that shows empirical performance that matches the anticipated pre-log factor 
of $1-\frac{S}{N}$.

The results of this work are only a first step towards the
understanding of reliable communication over sparse channels.
Important open questions concern rigorous analyses of the cases that
i) the inactive channel taps are not exactly zero-valued,
ii) the channel has \emph{at most} (rather than exactly) $S$ active taps,
iii) the receiver does not know the channel statistics, 
iv) the channel taps are correlated within and/or across blocks, and/or
v) the channel taps are non-Gaussian.

\appendices

\section{Proof of \lemref{optd}}		 \label{app:optd}
 
\begin{IEEEproof}
The maximum a posteriori (MAP) codeword estimate is defined as 
\begin{eqnarray}
  \hvec{x}\da\map
  &=& \argmax_{\vec{x}\da\in\code} 
  	p\big( \vec{x}\da \biggiv \{\vec{y}\of{k}\da\}_{k=1}^K, 
		\{\vec{y}\of{k}\pl\}_{k=1}^K, 
  		\vec{x}\pl \big) 			\label{eq:dmap0}\\
  &=& \argmax_{\vec{x}\da\in\code} 
  	p\big( \{\vec{y}\of{k}\da\}_{k=1}^K \biggiv 
		\vec{x}\da, \{\vec{y}\of{k}\pl\}_{k=1}^K, \vec{x}\pl \big) \,
	p(\vec{x}\da)					\label{eq:dmap1}
\end{eqnarray}
where \eqref{dmap1} results after applying Bayes rule and simplifying.
Assuming that codewords are uniformly distributed over $\code$, 
the MAP codeword estimate reduces to the maximum likelihood estimate 
\begin{eqnarray}
  \hvec{x}\da\ml
  &=& \argmax_{\vec{x}\da\in\code} 
  	p\big( \{\vec{y}\of{k}\da\}_{k=1}^K \biggiv 
		\{\vec{x}\da\of{k}\}_{k=1}^K, 
		\{\vec{y}\of{k}\pl\}_{k=1}^K, \vec{x}\pl \big) \\
  &=& \argmax_{\vec{x}\da\in\code} 
        \prod_{k=1}^K\sum_{i=1}^M 
	\Pr\{\mc{L}\of{k}=\mc{L}_i \giv 
		\vec{x}\da\of{k},\vec{y}\pl\of{k},\vec{x}\pl\}
	\nonumber\\&&\mbox{}\times
  	\int_{\vec{h}\of{k}\nz}
  	p\big( \vec{y}\of{k}\da \giv 
		\vec{x}\da\of{k}, \vec{y}\pl\of{k}, \vec{x}\pl,
		\vec{h}\of{k}\nz, \mc{L}\of{k}=\mc{L}_i\big) 
	p\big( \vec{h}\of{k}\nz \giv \vec{x}\da\of{k}, 
		\vec{y}\of{k}\pl, \vec{x}\pl,\mc{L}\of{k}=\mc{L}_i \big) 
							\label{eq:dml1}\\
  &=& \argmax_{\vec{x}\da\in\code} 
        \prod_{k=1}^K\sum_{i=1}^M 
	\Pr\{\mc{L}\of{k}=\mc{L}_i \giv \vec{y}\pl\of{k},\vec{x}\pl\}
	\nonumber\\&&\mbox{}\times
  	\int_{\vec{h}\of{k}\nz}
  	p\big( \vec{y}\of{k}\da \giv 
		\vec{x}\da\of{k}, 
		\vec{h}\of{k}\nz, \mc{L}\of{k}=\mc{L}_i\big) 
	p\big( \vec{h}\of{k}\nz \giv 
		\vec{y}\of{k}\pl, \vec{x}\pl,\mc{L}\of{k}=\mc{L}_i \big) ,
							\label{eq:dml2}
\end{eqnarray}
where the decoupling in \eqref{dml1} is due to independent fading and
noise across fading-blocks.
Recalling that, under the hypothesis $\mc{L}\of{k}=\mc{L}_i$, the
pilot observations become
\begin{eqnarray}
  \vec{y}\pl\of{k} 
  &=& \sqrt{\rho N}\Diag(\vec{x}\pl)\vec{F}\pli \vec{h}\nz\of{k}
  	+\vec{v}\pl\of{k},				\label{eq:ypli}
\end{eqnarray}
with $p(\vec{h}\of{k}\nz\giv\mc{L}\of{k}=\mc{L}_i) 
= \mc{CN}(\vec{h}\of{k}\nz;\vec{0},S^{-1}\vec{I})$,
the posterior $p\big( \vec{h}\of{k}\nz \giv 
\vec{y}\of{k}\pl, \vec{x}\pl,\mc{L}\of{k}=\mc{L}_i \big)$ is Gaussian. 
In particular,
\begin{eqnarray}
  p(\vec{h}\of{k}\nz \giv \vec{y}\of{k}\pl, \vec{x}\pl, \mc{L}\of{k}=\mc{L}_i )
  &=& \mc{CN}\big(\vec{h}\of{k}\nz;
  	\hvec{h}\of{k}\nzpi,\vec{\Sigma}\nzpi\big),	
							\label{eq:phnzpi}
\end{eqnarray}
where $\hvec{h}\of{k}\nzpi$ can be recognized as the $\mc{L}_i$-conditional 
pilot-aided MMSE estimate of $\vec{h}\of{k}\nz$ and 
$\vec{\Sigma}\nzpi$ as its error covariance: 
\begin{eqnarray}
  \hvec{h}\of{k}\nzpi
  &\defn& \E\{\vec{h}\of{k}\nz 
  	\giv \vec{y}\of{k}\pl, \vec{x}\pl, \mc{L}\of{k}=\mc{L}_i\}\\
  \vec{\Sigma}\nzpi
  &\defn& \cov\{\vec{h}\of{k}\nz
  	\giv \vec{y}\of{k}\pl, \vec{x}\pl, \mc{L}\of{k}=\mc{L}_i\} .
\end{eqnarray}
Due to the linear Gaussian model \eqref{ypli}, the MMSE estimate 
$\hvec{h}\of{k}\nzpi$ is a linear function of $\vec{y}\of{k}\pl$:
\begin{eqnarray}
  \hvec{h}\of{k}\nzpi
  &=& \E\big\{ \vec{h}\of{k}\nz \vec{y}\ofH{k}\pl 
  	\biggiv \vec{x}\pl, \mc{L}\of{k} = \mc{L}_i \big\} 
     	\E\big\{ \vec{y}\of{k}\pl \vec{y}\ofH{k}\pl 
     	\biggiv \vec{x}\pl, \mc{L}\of{k} = \mc{L}_i \big\}^{-1} 
	\vec{y}\of{k}\pl\\
  &=& \textstyle \frac{\sqrt{\rho N}}{S}\vec{F}\pli\herm
  	\Diag(\vec{x}\pl^*) \big( \rho\frac{N}{S}\Diag(\vec{x}\pl)
	\vec{F}\pli\vec{F}\pli\herm 
	\Diag(\vec{x}\pl^*) + \vec{I}\big)^{-1} \vec{y}\of{k}\pl \\
  &=& \textstyle \sqrt{\frac{\rho}{N}}\vec{F}\pli\herm
  	\big( \rho\vec{F}\pli\vec{F}\pli\herm 
	+ \frac{S}{N}\vec{I}\big)^{-1} \Diag(\vec{x}\pl^*)
	\vec{y}\of{k}\pl ,			\label{eq:hnzpi1}
\end{eqnarray}
where, for \eqref{hnzpi1}, we exploited the fact that $\vec{x}\pl$ has
constant-modulus elements.
Similarly,
\begin{eqnarray}
  \vec{\Sigma}\nzpi
  &=& \E\big\{ \vec{h}\of{k}\nz \vec{h}\ofH{k}\nz 
  	\biggiv \mc{L}\of{k} = \mc{L}_i \big\} 
	- \E\big\{ \vec{h}\of{k}\nz \vec{y}\ofH{k}\pl 
  	\biggiv \vec{x}\pl, \mc{L}\of{k} = \mc{L}_i \big\} 
	\nonumber\\&&\mbox{}\textstyle\times
	\E\big\{ \vec{y}\of{k}\pl \vec{y}\ofH{k}\pl 
  	\biggiv \vec{x}\pl, \mc{L}\of{k} = \mc{L}_i \big\}^{-1}
	\E\big\{ \vec{y}\of{k}\pl \vec{h}\ofH{k}\nz 
  	\biggiv \vec{x}\pl, \mc{L}\of{k} = \mc{L}_i \big\}  \\
  &=& \textstyle \frac{1}{S}\vec{I} 
  	- \frac{\rho N}{S^2} \vec{F}\pli\herm \Diag(\vec{x}\pl^*) 
	\big( \frac{\rho N}{S}\Diag(\vec{x}\pl)
        	\vec{F}\pli\vec{F}\pli\herm  
        	\Diag(\vec{x}\pl^*) + \vec{I}\big)^{-1}
	\Diag(\vec{x}\pl) \vec{F}\pli\\
  &=& \textstyle \frac{1}{S}\Big(\vec{I} - \vec{F}\pli\herm
	\big( \vec{F}\pli\vec{F}\pli\herm 
        	+ \frac{S}{\rho N}\vec{I}\big)^{-1}
	\vec{F}\pli \Big) .
\end{eqnarray}

Finally, since both pdfs in \eqref{dml2} are Gaussian, the integral can be 
evaluated in closed form, reducing to (see, e.g., \cite{Reader:EUSIPCO:96}) 
\begin{eqnarray}
  \lefteqn{ 
  \int_{\vec{h}\of{k}\nz}
  p\big( \vec{y}\of{k}\da \giv 
	\vec{x}\da\of{k}, 
	\vec{h}\of{k}\nz, \mc{L}\of{k}=\mc{L}_i\big) 
  p\big( \vec{h}\of{k}\nz \giv 
	\vec{y}\of{k}\pl, \vec{x}\pl,\mc{L}\of{k}=\mc{L}_i \big)
  }\nonumber\\
  &=&
  C\det\Big(\rho N \vec{F}\di\herm
  	\Diag(\vec{x}\da\of{k}\odot\vec{x}\da\ofc{k})\vec{F}\di
  	+ \vec{\Sigma}\nzpi^{-1} \Big)^{-1}
  \nonumber\\&&\mbox{}\textstyle\times
  \exp\Big(-\big\|\vec{y}\of{k}\da-\sqrt{\rho N}\Diag(\vec{x}\of{k}\da)
  	\vec{F}\di \hvec{h}\of{k}\nzi(\vec{x}\of{k}\da) \big\|^2  
	- \big\|\hvec{h}\of{k}\nzi(\vec{x}\of{k}\da) - \hvec{h}\of{k}\nzpi
	\big\|^2_{\vec{\Sigma}^{-1}\nzpi} \Big),
							\label{eq:dml_int}
\end{eqnarray}
where $C$ does not depend on $\vec{x}\da$, and where
$\hvec{h}\of{k}\nzi(\vec{x}\of{k}\da)$ 
denotes the MMSE estimate of $\vec{h}\of{k}\nz$ conditioned on the 
data hypothesis $\vec{x}\of{k}\da$ and based on the pilot-aided prior
statistics \eqref{phnzpi}:
\begin{eqnarray}
  \hvec{h}\of{k}\nzi(\vec{x}\of{k}\da)
  &=& \hvec{h}\of{k}\nzpi + \sqrt{\rho N}\vec{\Sigma}\nzpi 
  	\vec{F}\di\herm\Diag(\vec{x}\ofc{k}\da)
	\big(\rho N\Diag(\vec{x}\of{k}\da)\vec{F}\di
		\vec{\Sigma}\nzpi 
		\vec{F}\di\herm\Diag(\vec{x}\ofc{k}\da)
	+ \vec{I} \big)^{-1}
	\nonumber\\&&\mbox{}\times
	\big(\vec{y}\of{k}\da 
		- \sqrt{\rho N}\Diag(\vec{x}\of{k}\da)\vec{F}\di 
		\hvec{h}\of{k}\nzpi \big) .
\end{eqnarray}
\end{IEEEproof}

\section*{Acknowledgment}
The authors thank the anonymous reviewers for their insightful comments.

\begin{IEEEbiography}{Arun Pachai Kannu} 
received the M.S. and Ph.D. degrees in Electrical
Engineering from The Ohio State University in 2004 and 2007, respectively.
From 2007 to 2009, he was a Senior Engineer in Qualcomm Inc., San Diego, CA.
He is currently an Assistant Professor in the Department of Electrical
Engineering, Indian Institute of Technology, Madras.
\end{IEEEbiography}

\begin{IEEEbiography}{Philip Schniter} 
received the B.S. and M.S. degrees in Electrical and Computer Engineering 
from the University of Illinois at Urbana-Champaign in 1992 and 1993, 
respectively. 
From 1993 to 1996 he was employed by Tektronix Inc. in Beaverton, OR as a 
systems engineer, and 
in 2000, he received the Ph.D. degree in Electrical Engineering from 
Cornell University in Ithaca, NY. 
Subsequently, he joined the Department of Electrical and Computer Engineering 
at The Ohio State University in Columbus, OH, where he is now an Associate 
Professor and a member of the Information Processing Systems (IPS) Lab. 
In 2003, he received the National Science Foundation CAREER Award, and
in 2008-2009 he was a visiting professor at Eurecom (Sophia Antipolis, France) 
and Sup{\'e}lec (Gif-sur-Yvette, France).
\end{IEEEbiography}



\end{document}